\documentclass[fleqn,usenatbib]{mnras}

\usepackage{newtxtext,newtxmath}
\usepackage[T1]{fontenc}

\DeclareRobustCommand{\VAN}[3]{#2}
\let\VANthebibliography\thebibliography
\def\thebibliography{\DeclareRobustCommand{\VAN}[3]{##3}\VANthebibliography}

\usepackage{graphicx}
\usepackage{subcaption}
\usepackage{amsmath}
\usepackage{xcolor}
\usepackage{orcidlink}

\newcommand{\Rg}{R_{\rm g}}

\title[X-ray Comptonization in a tilted-disc Kerr black hole system I]{X-ray Comptonization in an Accreting Kerr Black Hole System with a Tilted Disc and Thermal Corona; A General-Relativistic Monte Carlo Simulation: I. Model, Radiative Transfer and Spectral-Timing Properties}

\author[Mainak Chatterjee \& Dipankar Bhattacharya]{
Mainak Chatterjee,$^{1}$\orcidlink{0009-0005-1036-4523}\thanks{mainak.chatterjee\_phd23@ashoka.edu.in; mainak216@gmail.com}
Dipankar Bhattacharya$^{1}$\orcidlink{0000-0003-3352-3142}\thanks{dipankar.bhattacharya@ashoka.edu.in}
\\
$^{1}$Department of Physics, Ashoka University, Sonipat, Haryana 131029, India
}

\date{Accepted XXX. Received YYY; in original form ZZZ}
\pubyear{2026}

\begin{document}
\label{firstpage}
\pagerange{\pageref{firstpage}--\pageref{lastpage}}
\maketitle

\begin{abstract}
Quasi-periodic oscillations (QPOs) in the X-ray emission of accreting black hole binaries are commonly attributed to the relativistic Lense--Thirring precession of the inner accretion flow, in which a hot inner flow, tilted with respect to the black hole spin axis, precesses as a solid body while a cooler outer disc remains at its own, different tilt -- a warped, Bardeen--Petterson-type geometry. Testing this picture requires a radiative-transfer model that self-consistently follows photons through such a system: from emission near the innermost stable circular orbit (ISCO), through Compton scattering in a hot corona, to possible reflection off the misaligned outer disc, all under the full effects of light bending in Kerr spacetime. In this first paper of a two-part series, we present a general-relativistic Monte Carlo (GRMC) radiative-transfer model of a misaligned, warped, two-zone accretion system around a Kerr black hole. We describe the disc--corona geometry, the seed-photon prescription, the photon-transport engine -- including the geodesic integration scheme, the coronal Compton-scattering module and the outer-disc reflection treatment -- and the observer pipeline used to construct energy spectra, photon path-length and arrival-time distributions. We present the resulting energy spectra, light curves and spectral lags as functions of the black hole mass, corona size, outer-disc geometry, reflection albedo, photon-injection radius and observer inclination, and show how gravitational light bending systematically lengthens the path-length and arrival-time distributions of the escaping radiation. This model and its spectral-timing predictions form the foundation for Paper II, which extends the framework to track the full photon polarization state and presents the resulting energy- and precession-phase-resolved polarimetric signatures.
\end{abstract}


\section{Introduction}
\label{sec:intro}

Most black hole X-ray binaries (BHXRBs) are transients. During an outburst they
move through a sequence of spectral states \citep{BelloniMotta2016}, and in the
hard and hard-intermediate states the power spectrum very often shows a narrow
low-frequency quasi-periodic oscillation (LFQPO) somewhere between about 0.1 and
20\,Hz \citep{IngramMotta2019}. These are sorted into types A, B and C by their
frequency, amplitude and coherence \citep{Casella2005}. Type-C QPOs are the
strongest and the most common, and they are the ones we are concerned with.

What produces the type-C QPO is still not settled. One family of models is
geometric: the emitting region changes its orientation relative to the observer
periodically, and the flux follows. The relativistic precession model ties the
QPO frequency to Lense--Thirring (LT) precession \citep{LenseThirring1918} of
matter close to the black hole \citep{StellaVietri1998, StellaVietriMorsink1999}.
For BHXRBs the best-developed version has a radially extended hot inner flow,
sitting inside a truncated thin disc, that precesses as a solid body
\citep{IngramDoneFragile2009, IngramDone2011}; precessing-ring variants also
exist \citep{SchnittmanHomanMiller2006, VeledinaPoutanenIngram2013}. The other
family treats the QPO as an intrinsic oscillation of the flow, for example a
Comptonizing corona coupled to the disc through feedback
\citep{Karpouzas2020, Bellavita2022, Mendez2022}, or as jet precession
\citep{Ma2021}.

The precession picture has real observational support. The centroid energy of the
iron line shifts with QPO phase in H1743$-$322, as expected if a precessing flow
illuminates different parts of the disc \citep{Ingram2016}. The photon index
oscillates with QPO phase \citep{IngramvdKlis2015, Ingram2016}, and both the QPO
amplitude and the phase lags depend on inclination in ways that relativistic
effects on the flux along different sightlines can account for
\citep{Heil2015, Motta2015, vandenEijnden2017}. There are also difficulties. The
very small truncation radius found in GRS\,1915+105, for instance, does not sit
comfortably with what the model expects \citep{Nathan2022}.

Any precession model needs the angular momentum of the flow to be misaligned with
the black hole spin, and that is not exotic: the supernova that forms the black
hole can tilt the orbit relative to the spin, and population synthesis suggests
that most Galactic BHXRBs end up within $\sim\!10^{\circ}$ of alignment, with a
tail reaching much larger angles \citep{Fragos2010}. Such misalignment has been
measured directly in MAXI\,J1820+070 \citep{Poutanen2022}, and a swinging jet in
V404\,Cyg has similarly been read as evidence for a precessing, tilted flow
\citep{MillerJones2019}.

A tilted disc does not stay flat. Differential LT precession warps it, and
viscosity drives the inner region into the equatorial plane of the black hole
while the outer region keeps its original tilt: the Bardeen--Petterson effect
\citep{BardeenPetterson1975}. Thin discs can respond dramatically to this
warping, even tearing into differentially precessing rings at large tilt
\citep{Liska2019, Liska2021, Nixon2012}; hot, geometrically thick flows behave
differently, and tilted GRMHD simulations show them precessing as a single rigid
unit instead \citep{Fragile2007, Liska2018}.

A natural starting point for the radiative signature of such a warped system
is the ray-tracing calculation of \citet{Abarr2020}. They take a geometrically
thin disc that extends from the ISCO to $100\,\Rg$ but is broken into two
planes: inside a Bardeen--Petterson radius of $8\,\Rg$ it is aligned with the
black hole spin, and outside it is tilted by $15^{\circ}$. They compute the
energy spectrum and polarization of the thermal emission, including photons
that return to the disc and reflect off both segments. An inner disc aligned
with the spin is not, however, the only possible outcome. In the viscous
regime, the inner disc may fail to align at all for a reasonable range of
parameters \citep{Banerjee2019a}, and the black hole itself can take
significantly longer to align with the outer disc than earlier estimates
suggested \citep{Martin2007, Banerjee2019b}. Observations also point to
a tilted inner disc in some sources: the behaviour of the iron line in
H1743$-$322 has been read as indicating a tilted inner disc
\citep{Ingram2016}, and in XTE\,J1550$-$564 the inclination inferred from
reflection ($\sim\!40^{\circ}$) is far below the orbital inclination
($\sim\!75^{\circ}$), with a misaligned inner disc among the explanations
discussed \citep{Connors2019}. A hot inner flow that is
itself tilted from the spin axis and precesses as a unit, surrounded by a
cooler, misaligned outer disc that stays put at a different tilt, is therefore
a sensible configuration to study, and it is the one we adopt.

Beyond timing and spectroscopy, X-ray polarimetry offers a further, largely
independent diagnostic of this geometry, since the polarization degree and
angle of scattered radiation depend sensitively on the scattering geometry and
on general-relativistic light bending. Modelling this polarimetric signature
self-consistently, however, requires the same underlying photon-transport
machinery developed here, extended to track the full Stokes vector of each
photon; we defer that extension to a
companion paper (Paper II).

Timing gives another, largely independent handle on this geometry.
Spectral lags between the coronal continuum and reflected disc emission
were first robustly detected in the narrow-line Seyfert 1 galaxy 1H0707$-$495
\citep{Fabian2009}, and their flux dependence has since been used to probe
changes in coronal geometry \citep{Kara2013}. Type-C QPO phase lags depend on
inclination \citep{vandenEijnden2017}, and are usually attributed to spectral
evolution over the precession cycle or to feedback within the Comptonizing flow
itself \citep{IngramvdKlis2015, Karpouzas2020, Bellavita2022}. In
Swift\,J1727.8$-$1613 the soft lags of $\gtrsim 10$\,ms cannot be pure
light-crossing delays if the corona lies in the disc plane \citep{Ingram2024}.
A Monte Carlo code that records photon path lengths and scattering histories
gives time delays directly, self-consistently with the energy spectrum, from
a single radiative-transfer calculation.

In this paper we build such a calculation. The system consists of a Kerr black
hole of mass $M_{\rm BH}$ and spin $a$, a hot inner flow tilted
by $\theta_{\rm in}$ from the spin axis, surrounded by a hot, spherical,
optically thin Comptonizing corona, and a stationary outer disc tilted by
$\theta_{\rm out}$ and truncated at its own Bardeen--Petterson warp
radius \citep{BardeenPetterson1975}. The observer is placed at inclination $i$
relative to the BH spin axis.

Photons follow null geodesics in Boyer--Lindquist coordinates, integrated
with an in-house, \texttt{numba}-compiled implementation of the DOP853
method, which makes the photon statistics needed for well-converged spectra
and lag distributions computationally tractable at this scale. In the corona,
photons Compton-scatter using a weighted Monte Carlo scheme with forced
collisions, and photons that reach the outer disc are
reflected. The thermal emission of the outer disc itself is computed
analytically from a Shakura--Sunyaev model \citep{ShakuraSunyaev1973} and
combined with the Monte Carlo output. The outputs are the energy-resolved
spectrum, photon path-length and light-curve distributions, and spectral
lags.

We investigate how gravitational light bending shapes the photon
path-length and arrival-time distributions of the escaping radiation, and
how the spectral lag between the direct coronal continuum and the
outer-disc-reflected photons depends on the black hole mass, corona size,
outer-disc geometry, reflection albedo and photon-injection radius. We examine how these parameters shape the resulting
spectral and timing behaviour, with the aim of turning observations into
constraints on them.

This model provides the common radiative-transfer foundation for a two-paper
series. Paper II extends the same photon-transport engine to track the full
photon polarization state, adding the Walker--Penrose parallel-transport
formalism and a polarized treatment of Compton and disc scattering, and uses
it to compute the energy- and precession-phase-resolved polarization degree
and angle of the composite system, which can be compared directly against IXPE
observations of BHXRBs such as Swift\,J1727.8$-$1613
\citep{Zhao2024} and against the enhanced sensitivity anticipated from the
upcoming eXTP mission.
\begin{figure}
    \centering
    \includegraphics[width=0.95\linewidth]{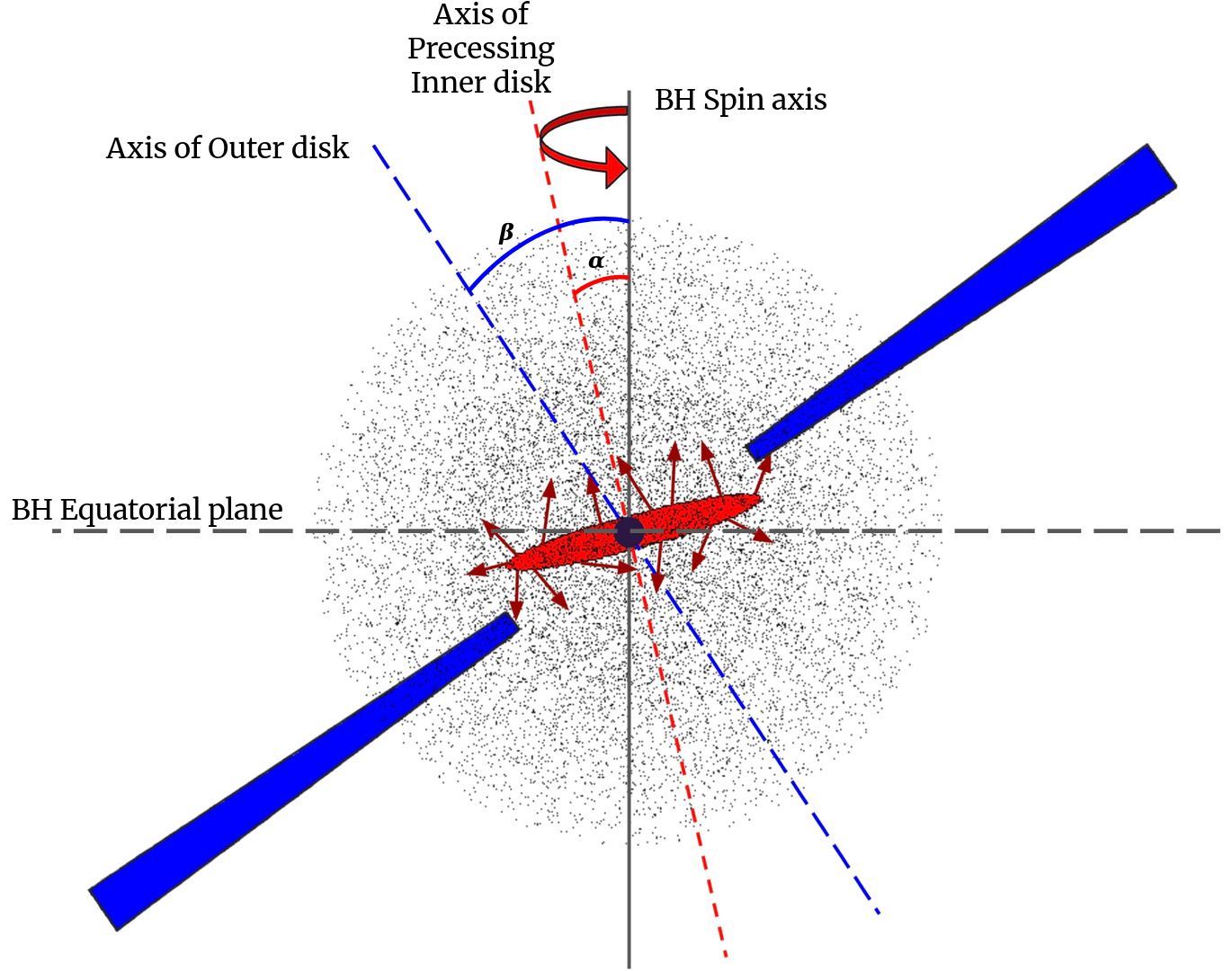}
    \caption{Schematic diagram of the proposed system. It comprises the BH--corona system, the precessing inner disc (red) and the stationary outer disc (blue), both tilted at different angles with respect to the BH spin axis. Here $\alpha$ and $\beta$ are $\theta_{\rm in}$ and $\theta_{\rm out}$ respectively.}
    \label{fig:system}
\end{figure}
\section{Model and Method}
\label{sec:system}
\subsection{System geometry}
\label{sec:disk_geometry}
The system (Fig.~\ref{fig:system}) is a two-zone accretion disc whose zones are tilted from the black hole spin axis $\hat{z}_{\rm BH}$ by $\theta_{\rm in}$ (inner disc, from the ISCO to the warp radius $r_{\rm warp}$) and $\theta_{\rm out}$ (outer disc, from $r_{\rm warp}$ to $r_{\rm out}$), motivated by the Bardeen--Petterson effect. Since the inner disc need not fully align \citep{Banerjee2019a}, we allow it a finite tilt, and adopt a sharp-transition approximation in which each zone is a locally flat disc with its own normal. A distant observer lies at inclination $i$ from the spin axis. We work in geometrized units ($G=c=1$), with lengths in units of $\Rg = GM_{\rm BH}/c^2$ and spin $a=J/M_{\rm BH}$. Positions are related to Boyer--Lindquist (BL) coordinates through the oblate-spheroidal Cartesian embedding
\begin{equation}
x = \sqrt{r^{2}+a^{2}}\,\sin\theta\cos\phi, \quad
y = \sqrt{r^{2}+a^{2}}\,\sin\theta\sin\phi, \quad
z = r\cos\theta.
\label{eq:oblate_cartesian}
\end{equation}

\subsection{Corona and seed photons}
\label{sec:corona}

The corona is a spherically symmetric region of hot electrons of radius $R_{\rm c}$ centred on the black hole, with density
\begin{equation}
n_{\rm e}(r)=n_{{\rm e}0}\exp(-r/R_{\rm c}),
\label{eq:elec_density}
\end{equation}
and temperature $T_{\rm e}$, in the spirit of the two-phase disc--corona picture \citep{HaardtMaraschi1991, ZdziarskiLubinskiSmith1999}. Seed photons are injected from the inner disc surface in a narrow annulus (radius $R_0$, width $\delta R_0$), with radii informed by fits of the Relativistic Precession Model to type-C QPO frequencies. Their energies are drawn from a Planck distribution at the local temperature
\begin{equation}
T(R) = \left[\frac{3 G M_{\rm BH} \dot{M}}{8\pi\sigma_{\rm SB} R^{3}}\right]^{1/4},
\label{eq:nt_temperature}
\end{equation}
the Shakura--Sunyaev profile \citep{ShakuraSunyaev1973} without the boundary term, which would artificially suppress emission near the ISCO. Emission directions are isotropic. Each photon also carries a polarization state, which plays no role in the energy-integrated observables here and is described in Paper II.

\subsection{Photon transport}
\label{sec:simulation}

The code is a weighted Monte Carlo engine. Each photon carries a 4-momentum $p^\mu$ and a weight $w$ (initially 1) encoding the probability of its history. Its guiding principle is that \emph{general relativity is used only to propagate photons between scattering events; all scattering physics is done in a local inertial frame with special relativity}. The zero-angular-momentum-observer (ZAMO) tetrad bridges the two regimes.

\paragraph{Geodesic propagation.}
Photons follow null geodesics of the Kerr metric in BL coordinates, integrated from Hamilton's equations (with $\mu=\cos\theta$ as the polar variable to avoid the pole singularity) using our own \texttt{numba}-compiled DOP853 integrator with $\mathrm{rtol}=10^{-9}$, $\mathrm{atol}=10^{-11}$. The energy at infinity $-p_t$ and axial angular momentum $p_\phi$ are conserved, and an auxiliary path-length variable is integrated alongside the geodesic. Events monitored along each trajectory are: horizon capture, ergosphere crossing and violation of the Hamiltonian constraint (photon killed); escape beyond an outer boundary; completion of the sampled free path (photon scatters); and crossing of the outer-disc plane within $[r_{\rm warp}, r_{\rm out}]$ (photon reflected). Redshifted photon energies and escape directions capture the effects of gravitational redshift, Doppler shift and frame dragging.

\paragraph{Coronal scattering.}
The free path is sampled between each pair of scatterings with the truncated exponential form \citep{PozdnyakovSobolSyunyaev1983}
\begin{equation}
\ell = -\bar\lambda\ln\!\left[1 - u\!\left(1 - e^{-(R_{\rm c} - r)/\bar\lambda}\right)\right],
\qquad \bar\lambda(r)=\frac{1}{n_{\rm e}(r)\sigma_{\rm T}},
\end{equation}
with $u$ a uniform deviate. At the end of the path the weight is split into an escaping fraction $w_{\rm esc}=e^{-(R_{\rm c}-r)/\bar\lambda}$ and a scattering fraction $w_{\rm scat}=1-w_{\rm esc}$, the latter updating $w\leftarrow w\,w_{\rm scat}$. For each scattering, the photon is transformed to the local Cartesian frame, a thermal electron is drawn from the Maxwell--J\"uttner distribution (isotropic direction), and the photon is boosted into the electron rest frame. There the outgoing direction is drawn from the unpolarized Thomson cross-section $\propto 1+\cos^2\Theta_{\rm scat}$ and the energy follows the Compton shift formula,
\begin{equation}
\label{eq:compton_energy_shift}
E'_{\rm scat} = \frac{E'_{\rm inc}}{1 + (E'_{\rm inc}/m_{\rm e}c^{2})(1 - \cos\Theta_{\rm scat})},
\end{equation}
after which the inverse boost returns the photon to the lab frame, yielding a net energy gain on average (Comptonization). A photon history ends on black-hole capture, escape, weight falling below $w_{\rm min}$ (treated as escaped), or exceeding the maximum number of scatterings (discarded).

\paragraph{Outer-disc reflection.}
When a photon crosses the outer-disc plane inside the physical annulus, it is treated as a single Compton scattering off cool electrons at rest: the outgoing direction is sampled from the Thomson cross-section, the energy is shifted by the Compton formula above (Equation \ref{eq:compton_energy_shift}), and the weight is multiplied by the albedo $\rho_{\rm alb}\in[0,1]$, the remainder being absorbed. The photon is then restarted from the disc surface, with safeguards preventing immediate re-triggering of the crossing event.

\subsection{Observer and observables}
\label{sec:observer_overview}

Escaping photons fall into two inner-disc populations: \emph{Component~A}, photons that never interact with the outer disc (including those scattered any number of times in the corona, and unscattered photons, which are still light-bent and redshifted), and \emph{Component~B}, photons that also reflect off the outer disc. Both are collected by a common \emph{patch observer} centred on the line of sight and referenced to the spin axis $\hat{n}_{\rm BH}$, the only axis fixed for the whole simulation about which the transport is axisymmetric. A photon with asymptotic direction $\hat{k}$ is accepted if
\begin{equation}
\cos(i+\Delta\theta)\le \hat{k}\cdot\hat{n}_{\rm BH}\le\cos(i-\Delta\theta),
\qquad |\varphi(\hat{k})|\le\Delta\phi,
\end{equation}
where $\varphi$ is the azimuth relative to the observer direction, and \textit{i} being the inclination of the observer with respect to the BH. Weighted sums of $w$, $wE$ and the arrival time $t$ are accumulated per energy bin and per component, and divided by the patch solid angle $\Omega_{\rm patch}=[\cos(i-\Delta\theta)-\cos(i+\Delta\theta)]\,2\Delta\phi$ and the bin width to give specific intensities $I_A(E)$ and $I_B(E)$,given by
\begin{equation}
I_X(E)=\frac{\sum_{j\in X,\,E} w_jE_j}{\Omega_{\rm patch}\,\Delta E},
\qquad X\in\{A,B\}
\end{equation} 
which add directly: $I_{\rm inner}=I_A+I_B$.

The \emph{spectral time lag} of component $X\in\{A,B\}$ is $\Delta t_X(E)=\langle t_X(E)\rangle_w - t_{\rm ref}$, where $\langle t_X\rangle_w$ is the weighted mean arrival time in the bin and $t_{\rm ref}$ is the broadband weighted mean over all accepted photons. The composite lag is the flux-weighted average,
\begin{equation}
\Delta t_{\rm inner}(E)=\frac{I_A\Delta t_A+I_B\Delta t_B}{I_A+I_B},
\end{equation}
with uncertainties propagated using the effective sample size $N_{\rm eff}=(\sum w)^2/\sum w^2$ \citep{Kish1965}.

The direct thermal emission of the outer disc (\emph{Component~C}) is not Monte Carlo sampled, but computed analytically from the Shakura--Sunyaev thin-disc solution \citep{ShakuraSunyaev1973, NovikovThorne1973} with
\begin{equation}
T_{\rm SS}(R)=T_0R^{-3/4}\left[1-\sqrt{R_{\rm outer,in}/R}\right]^{1/4},
\end{equation}
where $T_0=(3GM_{\rm BH}\dot M/8\pi\sigma_{\rm SB})^{1/4}$ and $R_{\rm outer,in}=r_{\rm warp}$. The boundary term is retained here, unlike for the inner ring. The flux at the observer is the Planck function integrated over the disc annuli, weighted by the projection factor $|\cos i_{\rm out}|$, with $i_{\rm out}$ the angle between $\hat n_{\rm out}$ and the line of sight, and binned onto the Monte Carlo energy grid.

\subsection{Normalization and composite spectrum}
\label{sec:fn2}

The Monte Carlo weights are dimensionless, so a single scale factor is needed to combine them with Component~C. Each launched photon carries the fraction $E_0/E_{\rm emit}$ of the ring luminosity, where $E_{\rm emit}=\sum_j E_{0,j}$ (as $w_0=1$) runs over all unique launched photons (including those later captured). With the ring luminosity $L_{\rm ring}=\tfrac{3}{2}GM_{\rm BH}\dot M\,(R_{\rm inner,in}^{-1}-R_{\rm inner,out}^{-1})$,
\begin{equation}
\texttt{mc\_scale}=\frac{L_{\rm ring}}{d^{2}\,E_{\rm emit}},
\qquad
I_{\rm tot}(E)=\texttt{mc\_scale}\,I_{\rm inner}(E)+F_E^{\rm out}(E).
\end{equation}
This preserves the absolute normalization implied by the emitted luminosity, so that photon capture, light bending, energy shifts, scattering and reflection are all propagated into the predicted flux, rather than being renormalized a posteriori to a bolometric flux. The outer disc contributes no Monte Carlo variance. The reported lag is that of the inner disc (A+B); the light-crossing delay of the outer disc's own emission is deferred to future work.

\section{Results and Analysis}
\label{sec:results_and_analysis}
\subsection{Fiducial model and parameter variations}
\label{sec:fiducial}

In this section we analyse the data generated by the simulation and present the main quantitative results. All simulations are built around a common fiducial model, whose parameters are described below. The system consists of a $10\,\mathrm{M_\odot}$ black hole with spin $a=0.9$, surrounded by a spherical corona of radius $R_{\rm c}=50\,\Rg$ whose electron-density profile (Eq.~\ref{eq:elec_density}) is normalized to $n_{{\rm e},0}=1\times10^{22}\,\mathrm{m^{-3}}$.

The inner and outer discs are tilted with respect to the BH spin axis (the $Z$ axis) by $\theta_{\rm in}=10^{\circ}$ and $\theta_{\rm out}=40^{\circ}$ respectively, with both tilts lying in the $Y$--$Z$ plane. The outer disc extends from $r_{\rm warp}=10\,\Rg$ to $r_{\rm out}=400\,\Rg$ and reflects photons with an albedo of $\rho_{\rm alb}=0.2$. Seed photons are emitted from an annular ring of radius $R_0=6.5\,\Rg$ and width $\delta R_0=0.1\,\Rg$ on the inner disc, and every simulation is started with $N_{\rm phot}=1\times10^{7}$ photons from this region. The accretion rate that sets the Shakura--Sunyaev temperature normalization is $\dot M = \dot{m} L_{\rm Edd}/(\eta c^2)$, where we adopt a radiative efficiency $\eta=0.1$ and an Eddington ratio $\dot{m} \equiv L/L_{\rm Edd}=0.1$, broadly representative of a black hole X-ray binary in a bright accretion state.

These fiducial values serve as the starting point for all runs. The observer is placed at an arbitrary reference distance of $d = 3.086\times10^{25}\,\mathrm{m}$ ($\simeq 1\,\mathrm{Gpc}$) from the source. This value can be changed freely: since the flux density scales as $d^{-2}$, varying $d$ only rescales the overall normalization of the energy (flux) spectrum, leaving its shape and the other plots unchanged. To show how the observables depend on the physical parameters of the system, we vary one parameter at a time while keeping the others at their fiducial values, unless stated otherwise. The parameter that is varied, together with its range, and the observer position and window ($\Delta\theta$, $\Delta\phi$) are given in the caption of each figure.
\subsection{Spectral properties of the model}
\begin{figure}
    \centering
    \includegraphics[width=\linewidth]{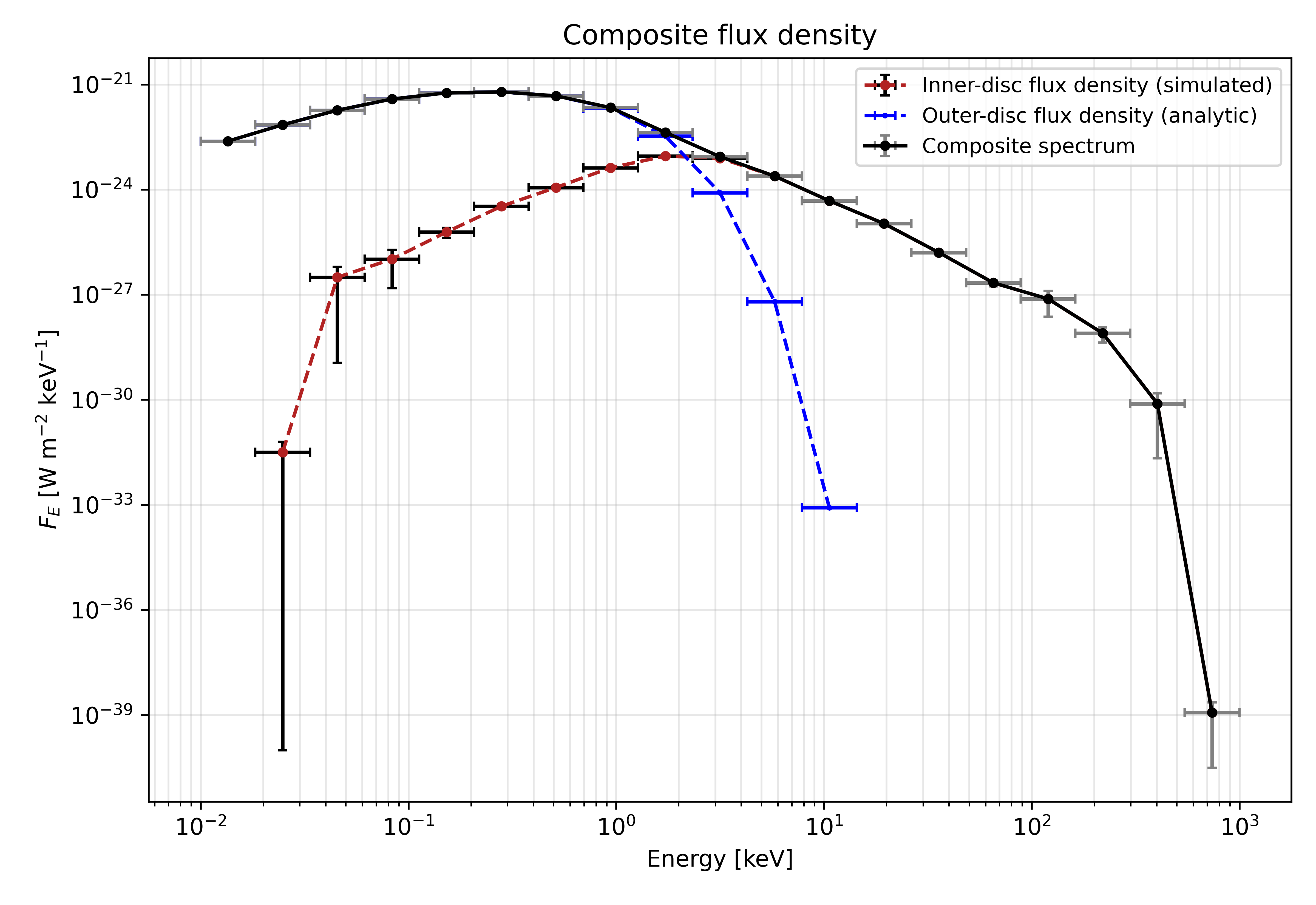}
    \caption{Composite spectrum of the system, as seen by an observer ($\Delta\theta=\pm3^{\circ}$, $\Delta\phi=\pm3^{\circ}$), at an inclination of $30^{\circ}$ with respect to the BH spin axis. The azimuthal angle is $30^{\circ}$ w.r.t. the global X-axis.}
    \label{fig:composit_spectra}
\end{figure}
\begin{figure}
    \centering
    \includegraphics[width=\linewidth]{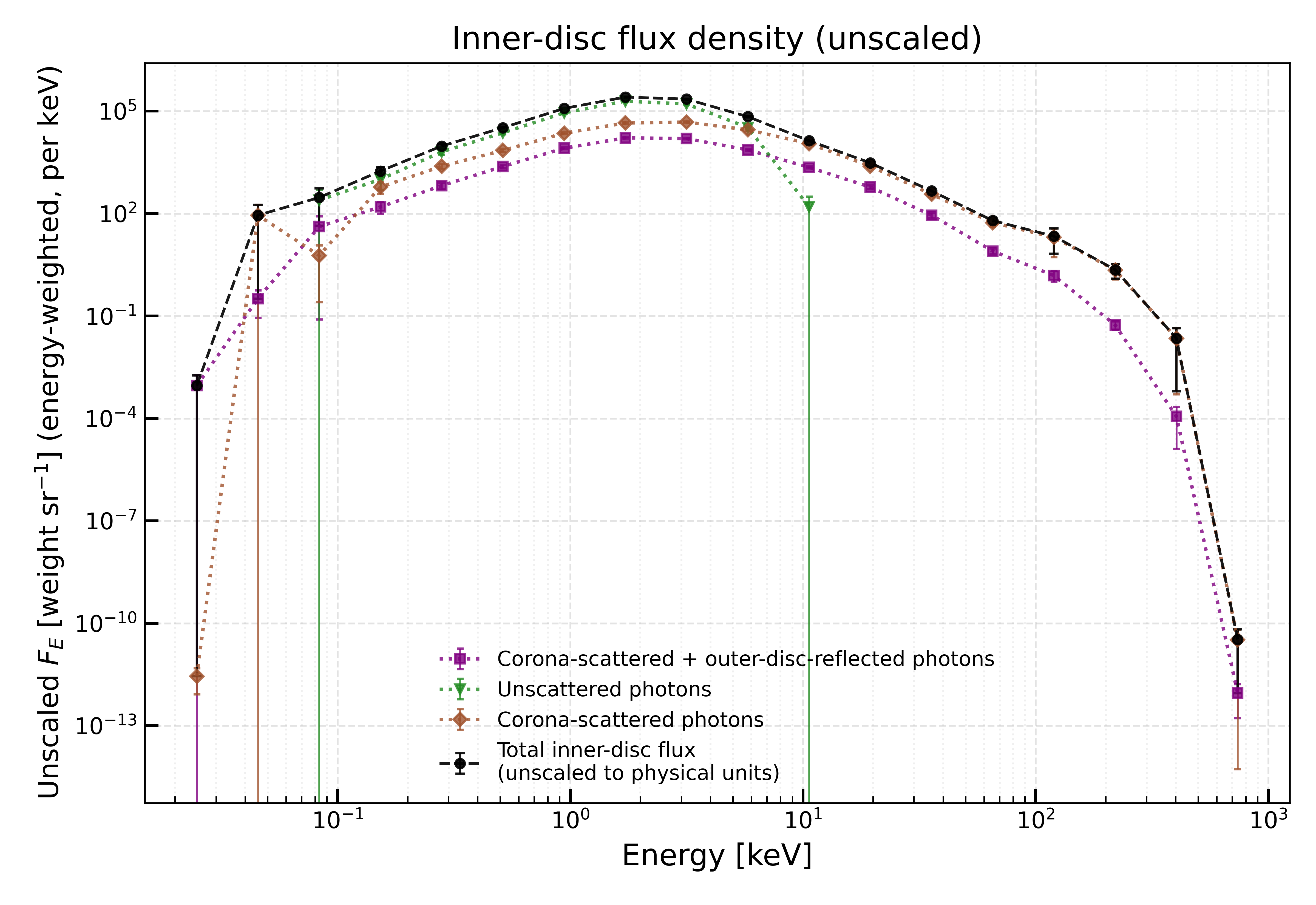}
    \caption{Raw composite spectrum from the inner disc and its individual components, as detected by an observer ($\Delta\theta=\pm3^{\circ}$, $\Delta\phi=\pm3^{\circ}$), at an inclination of $30^{\circ}$ with respect to the BH spin axis. The azimuthal angle is $30^{\circ}$ w.r.t. the global X-axis.}
    \label{fig:inner_spectra_combined}
\end{figure}
\begin{figure*}
    \centering
    \begin{subfigure}[b]{0.32\textwidth}
        \centering
        \includegraphics[width=\linewidth]{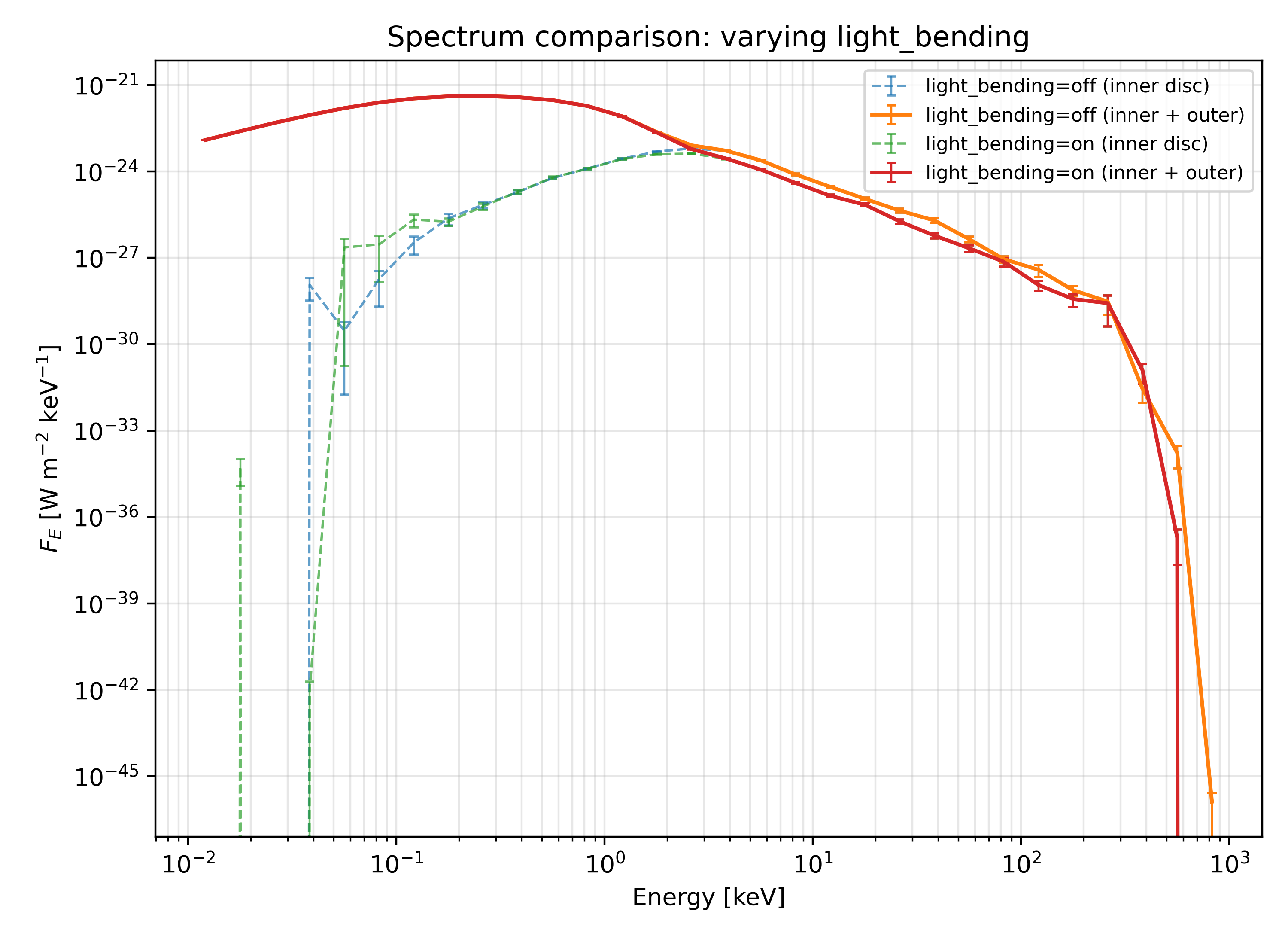}
        \caption{Spectra with light bending switched on and off.}
        \label{fig:light_bending_on_vs_off}
    \end{subfigure}
    \hfill
    \begin{subfigure}[b]{0.32\textwidth}
        \centering
        \includegraphics[width=\linewidth]{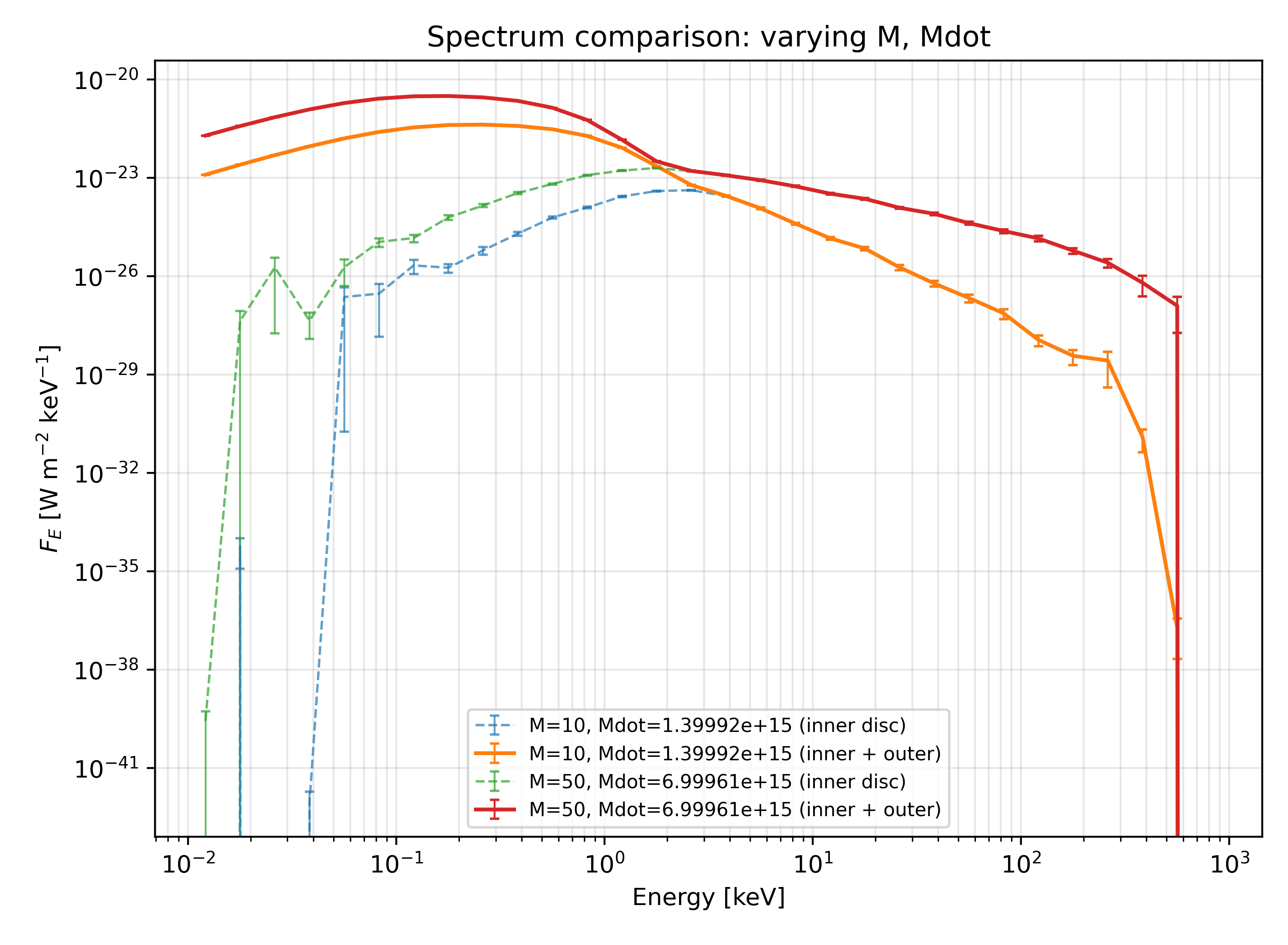}
        \caption{Spectra for varying black hole mass.}
        \label{fig:varying_mass}
    \end{subfigure}
    \hfill
    \begin{subfigure}[b]{0.32\textwidth}
        \centering
        \includegraphics[width=\linewidth]{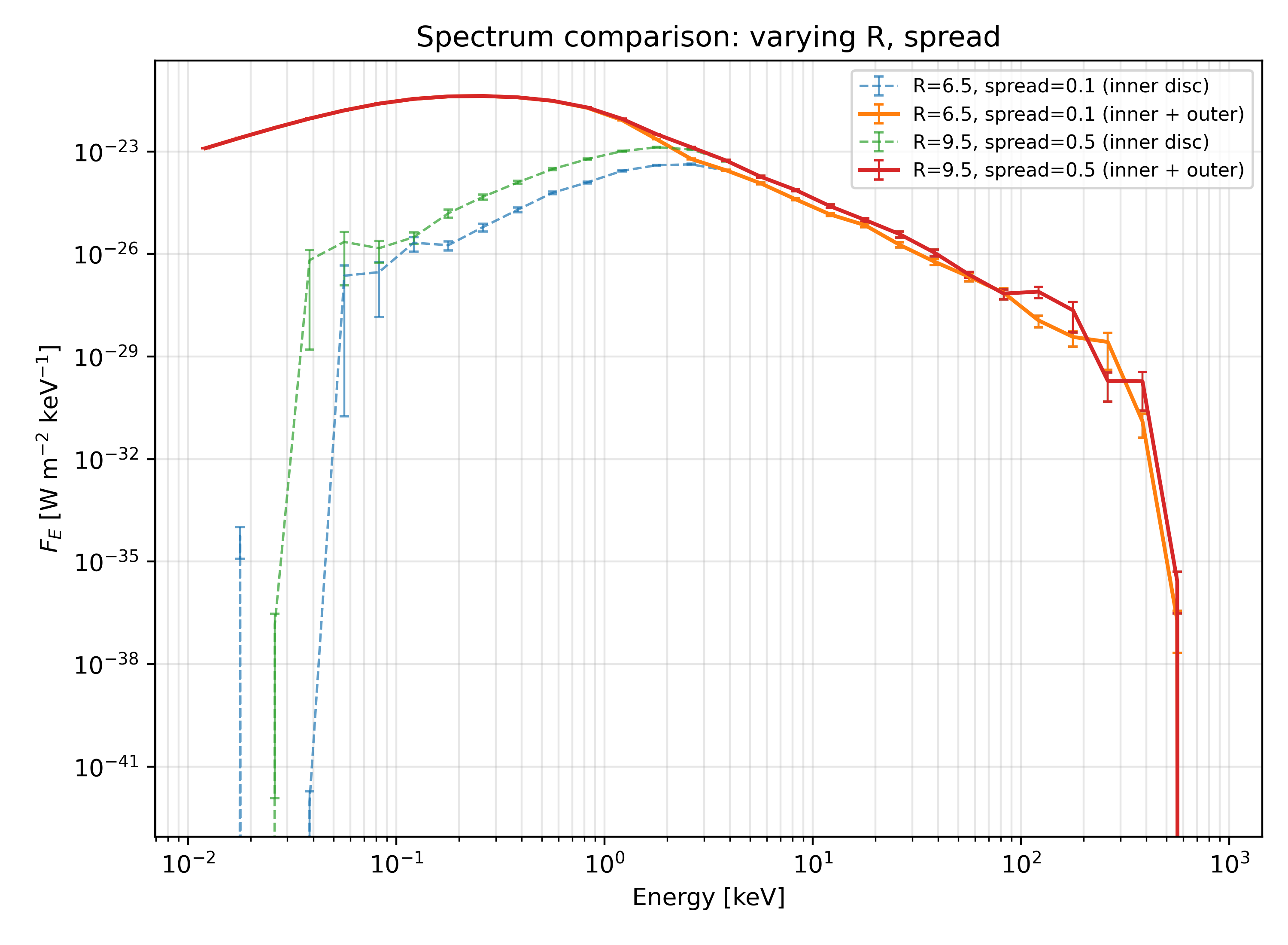}
        \caption{Spectra for varying photon-generation region.}
        \label{fig:varying_radius_and_width_of_photon_generation_region}
    \end{subfigure}
    \vspace{0.5cm}
    \begin{subfigure}[b]{0.32\textwidth}
        \centering
        \includegraphics[width=\linewidth]{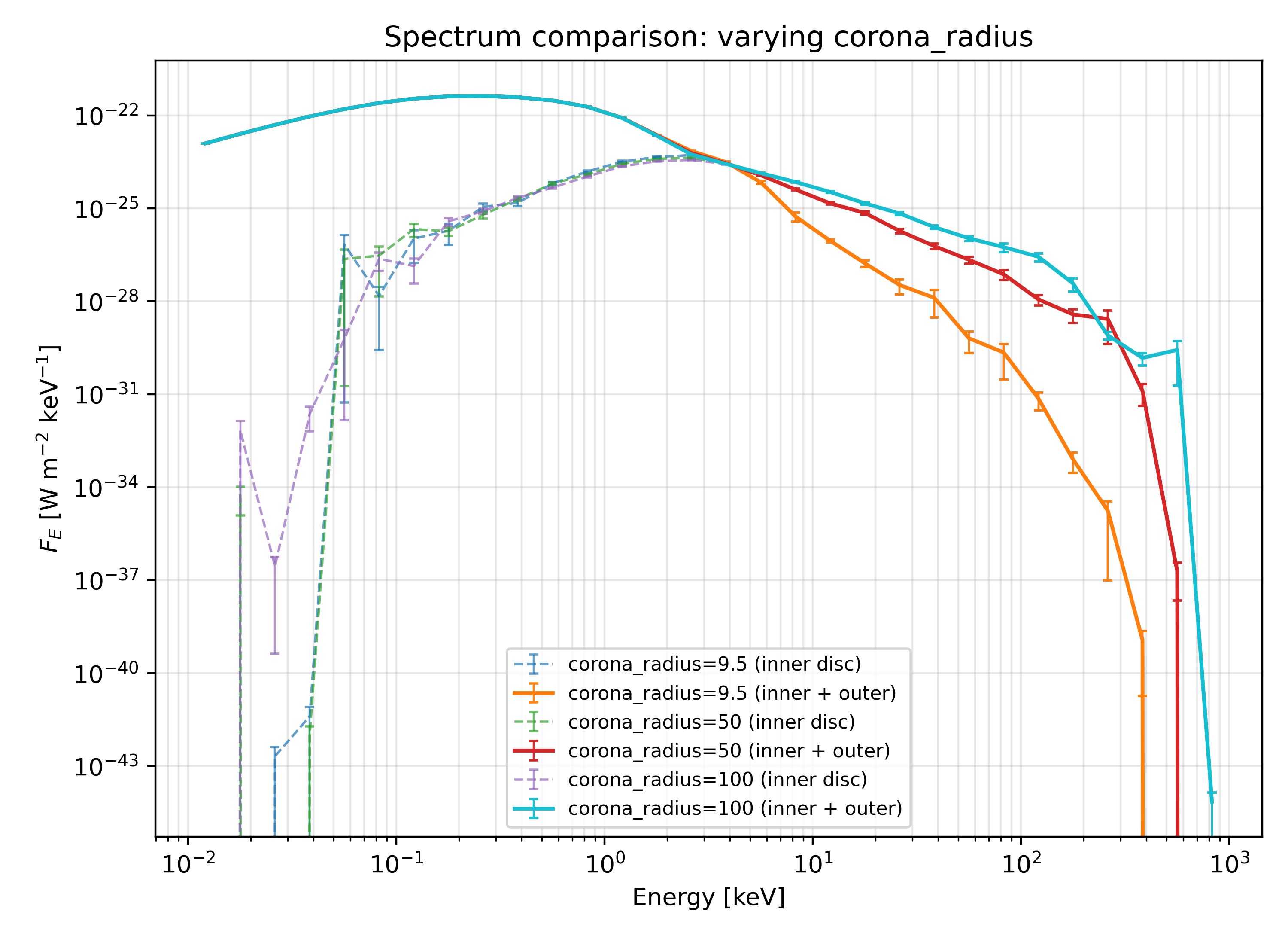}
        \caption{Effect of varying corona radius.}
        \label{fig:spectrum_on_changing_corona_radius}
    \end{subfigure}
    \hfill
    \begin{subfigure}[b]{0.32\textwidth}
        \centering
        \includegraphics[width=\linewidth]{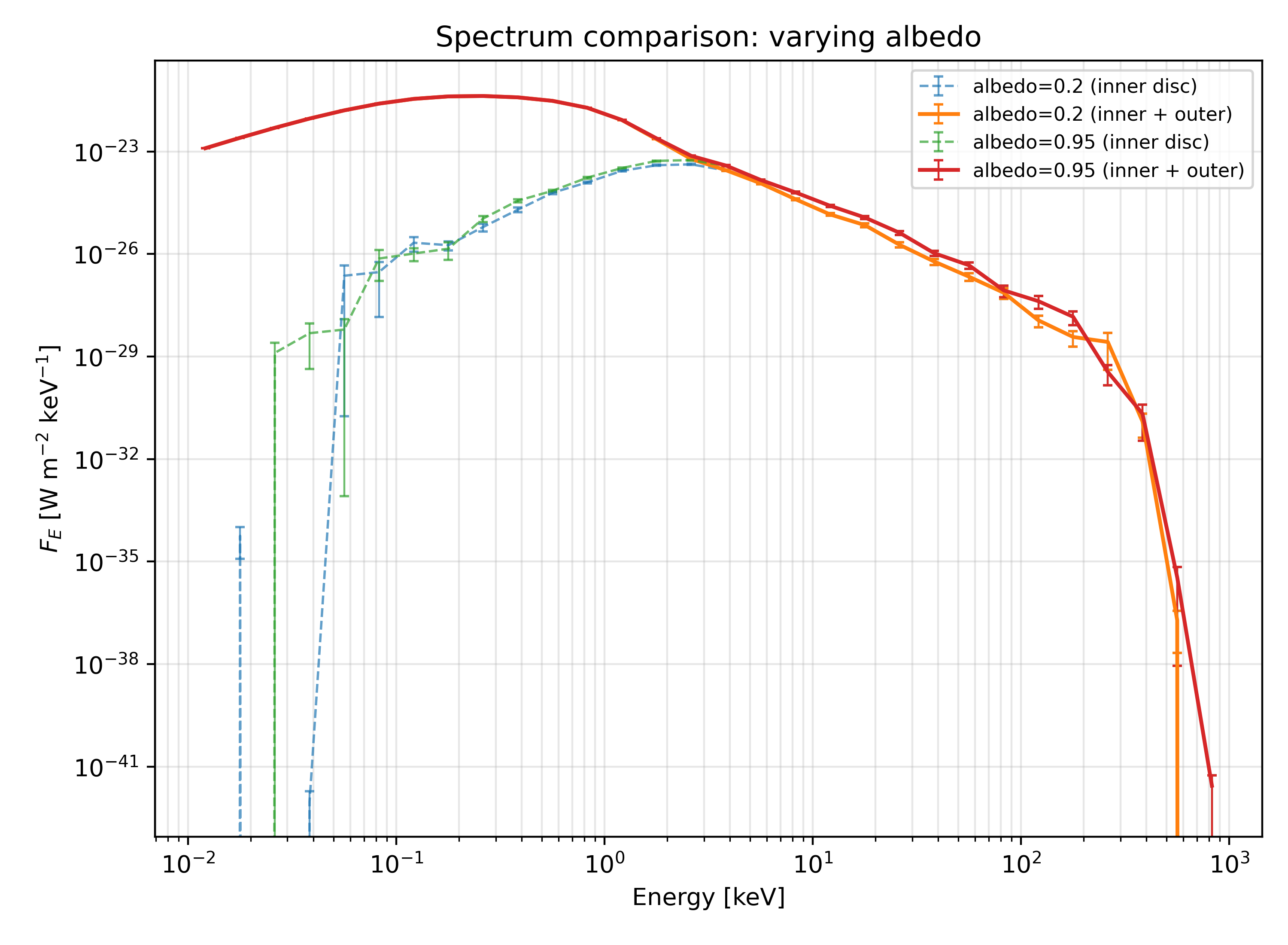}
        \caption{Effect of varying reflection albedo.}
        \label{fig:spectrum_on_changing_reflection_albeido}
    \end{subfigure}
    \hfill
    \begin{subfigure}[b]{0.32\textwidth}
        \centering
        \includegraphics[width=\linewidth]{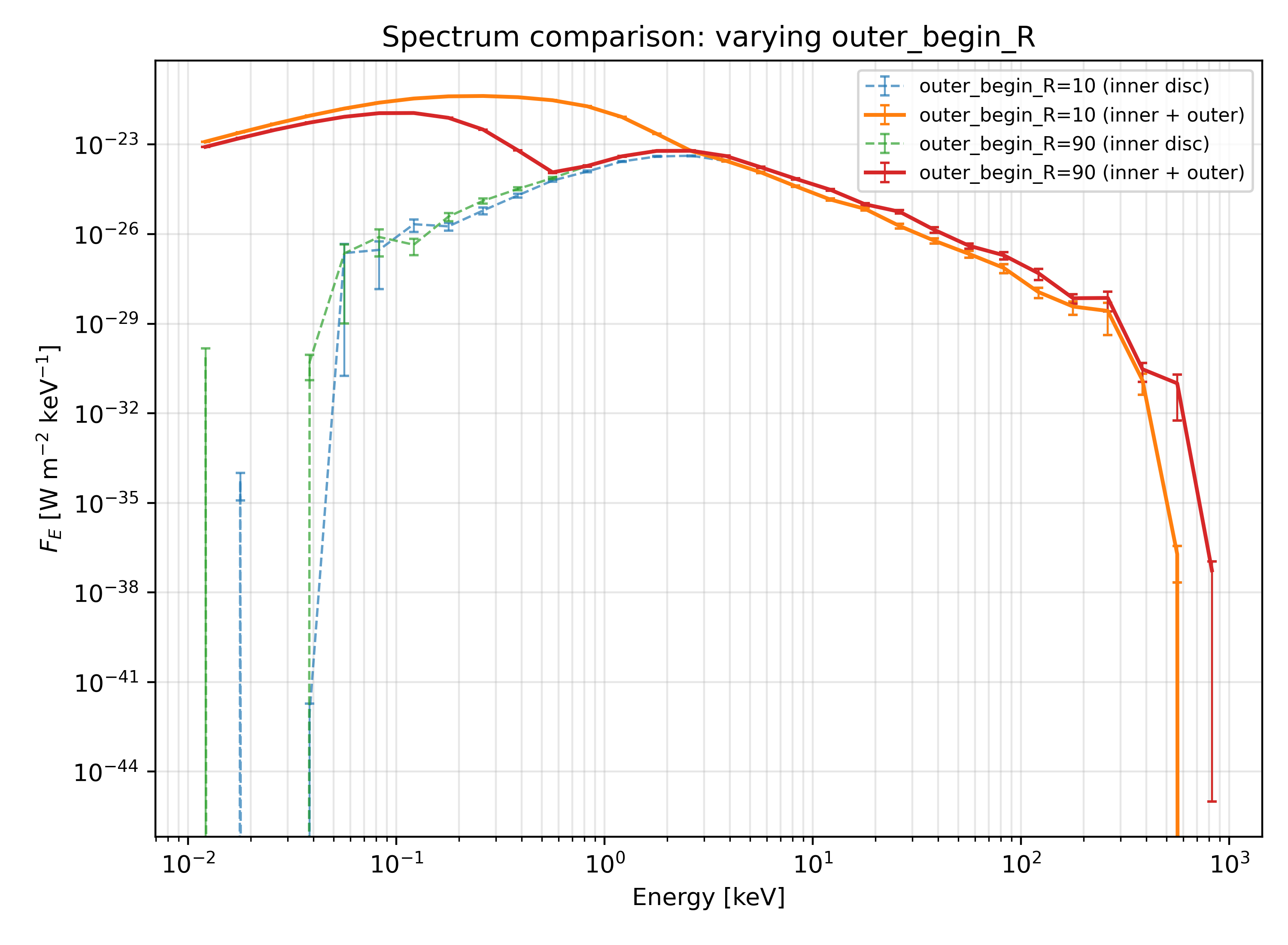}
        \caption{Effect of changing inner boundary of outer disc.}
        \label{fig:spectrum_on_changing_outer_disk_inner_radius}
    \end{subfigure}
    \caption{Comparison of simulated spectra under different physical and geometrical conditions. The individual panels show (a) the effect of light bending, (b) variation in black hole mass, (c) variation in the photon-generation radius, (d) variation in corona radius, (e) variation in reflection albedo, and (f) variation in the inner boundary of the outer disc, as observed through an observer window of $\Delta\theta=\pm3^{\circ}$, $\Delta\phi=\pm3^{\circ}$, the observer being positioned at an inclination of $30^{\circ}$ w.r.t. the BH spin axis, and having an azimuthal value of $90^{\circ}$ w.r.t. the global X-axis of the system. The R values mentioned in the legend are all in units of $R_g$.}
    \label{fig:combined_spectra}
\end{figure*}
\begin{figure}
    \centering
    \includegraphics[width=0.9\linewidth]{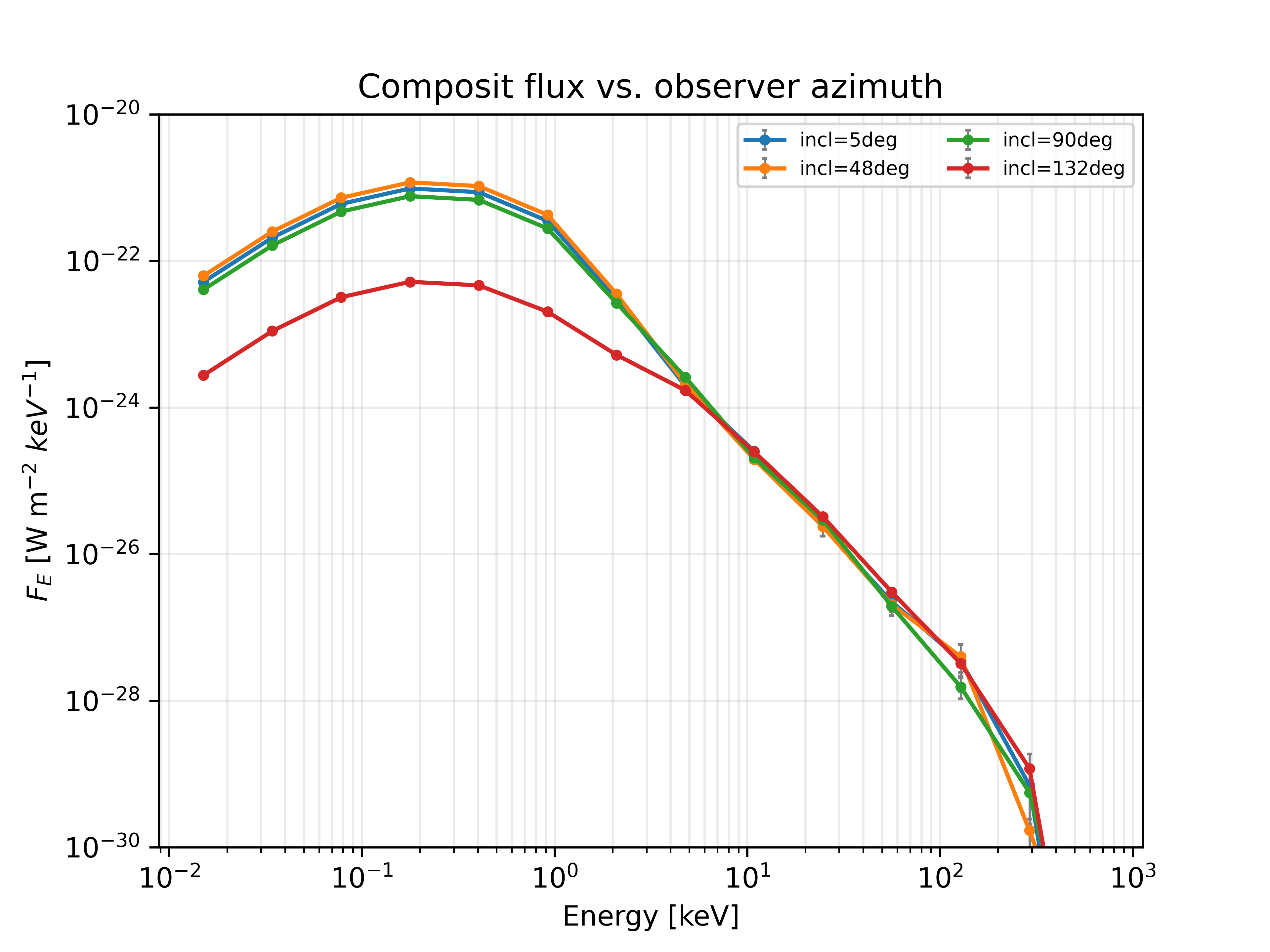}
    \caption{Composite spectrum of the system as observed at various inclination angles with respect to the BH spin axis, at a fixed azimuthal angle of $90^{\circ}$ ($\Delta\theta=\pm3^{\circ}$, $\Delta\phi=\pm3^{\circ}$).}
    \label{fig:spectrum_vs_inclination}
\end{figure}

Figure~\ref{fig:composit_spectra} shows the composite spectrum of the system as seen by an observer. The low-energy portion of the spectrum is dominated by the multicolour blackbody emission from the outer disc, while the high-energy portion is dominated by Compton up-scattered photons originating from the inner disc.
\\
The spectrum originating from the inner disc can be further decomposed into two components, as shown in Fig.~\ref{fig:inner_spectra_combined}. The lower-energy component is dominated by photons that have been reflected (scattered) at least once by the outer disc, while the higher-energy component is dominated by Compton up-scattered photons.
\\
The dependence of the composite spectrum on various physical system parameters is explored in Fig.~\ref{fig:combined_spectra}. We find that the spectral shape (and correspondingly the photon index) is the primary observable that vary with changes in the physical and initial parameters of the system.
\\
The dependence of the composite spectrum on observer inclination (with respect to the BH spin axis) is shown in Fig.~\ref{fig:spectrum_vs_inclination}. We find that the overall flux level varies with inclination.

\subsection{Inner-disc photon propagation}
\begin{figure*}
    \centering
    \begin{subfigure}[b]{0.45\linewidth}
        \centering
        \includegraphics[width=\linewidth]{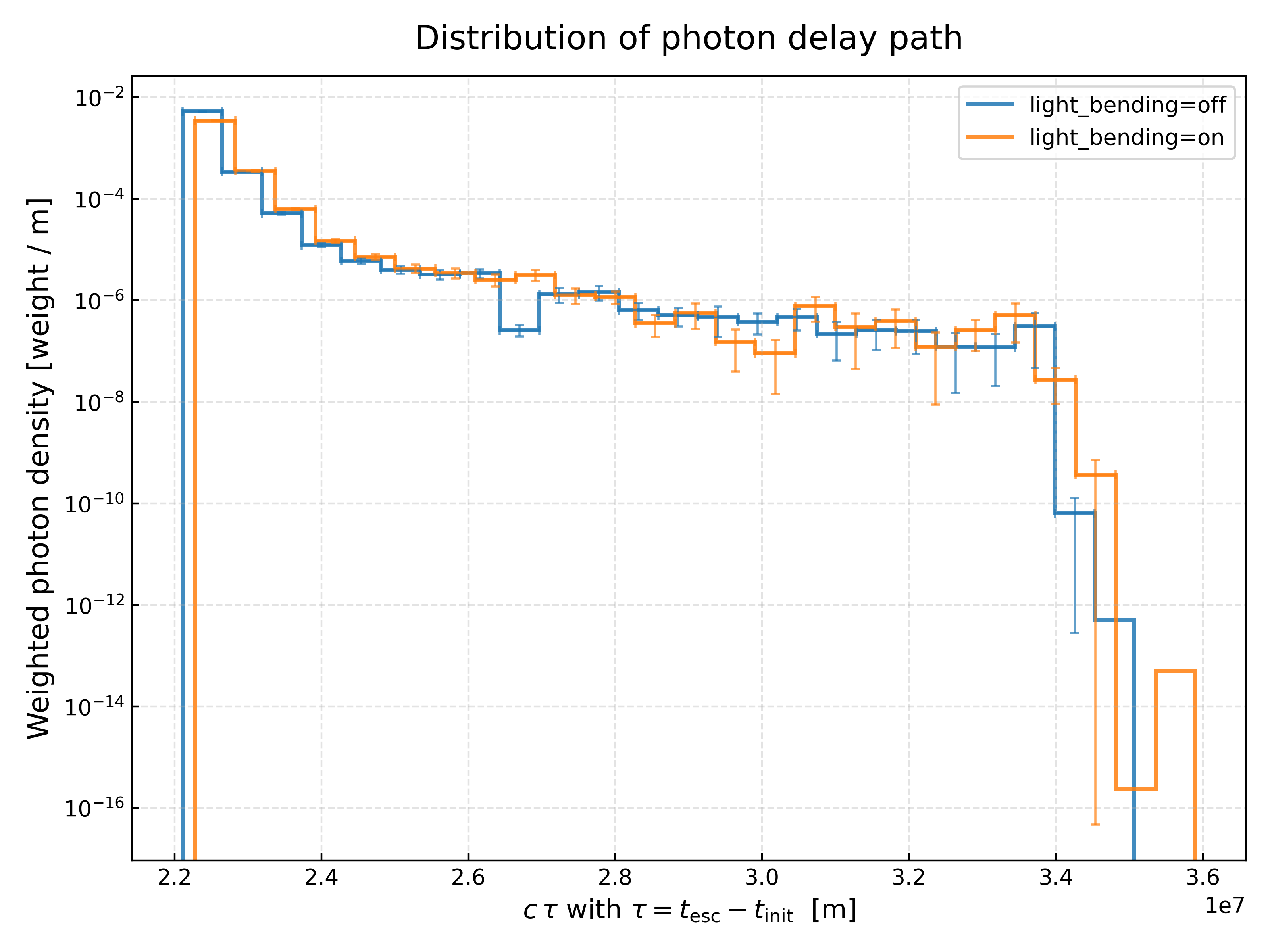}
        \caption{Light bending off vs on.}
        \label{fig:photon_path_length_comparison}
    \end{subfigure}
    \hfill
    \begin{subfigure}[b]{0.45\linewidth}
        \centering
        \includegraphics[width=\linewidth]{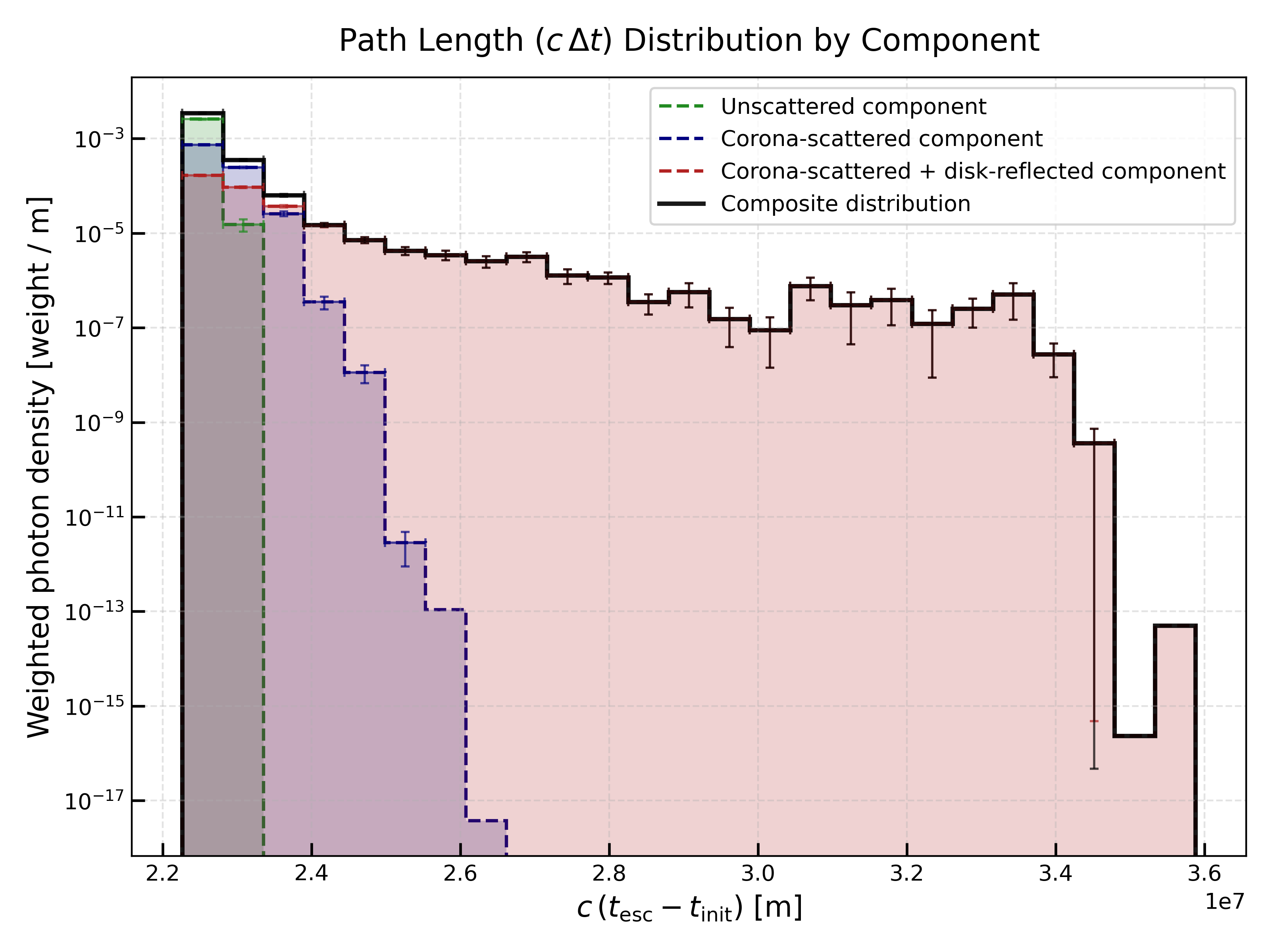}
        \caption{Photon path-length components.}
        \label{fig:photon_path_length_components}
    \end{subfigure}
    \caption{(a) Comparison of photon path-length distributions with light bending off and on, and (b) their decomposition with light bending on, as seen by an observer ($\Delta\theta=\pm3^{\circ}$, $\Delta\phi=\pm3^{\circ}$) at an inclination of $30^{\circ}$ with respect to the BH spin axis and an azimuthal angle of $90^{\circ}$ relative to the $X$-axis of the BH system.}
    \label{fig:photon_path_length}
\end{figure*}

We next examine the properties of photons emerging from the inner disc. Figure~\ref{fig:photon_path_length_comparison} shows the distribution of photon path lengths for an ensemble of photons generated during an instantaneous burst from the inner disc, for simulation runs both with and without light bending. The weighted photon density is found to remain elevated for longer path lengths when light bending is included, as evidenced by the more prominent tail of the distribution. We attribute this behaviour to light bending increasing the number of scattering events within the system as a result of curved photon trajectories, thereby increasing both the mean path length and the residence time of photons in the system. This is further corroborated by a delay in the arrival of the first photons in the light-bending case relative to the case without light bending.

Figure~\ref{fig:photon_path_length_components} decomposes this distribution into three classes: (a) direct photons that reach the observer without undergoing any scattering, (b) photons scattered within the corona, and (c) photons scattered within the corona and subsequently reflected by the outer disc.

\subsection{Light curves of instantaneous photon bursts}

\begin{figure*}
    \centering
    \begin{subfigure}[b]{0.45\linewidth}
        \centering
        \includegraphics[width=\linewidth]{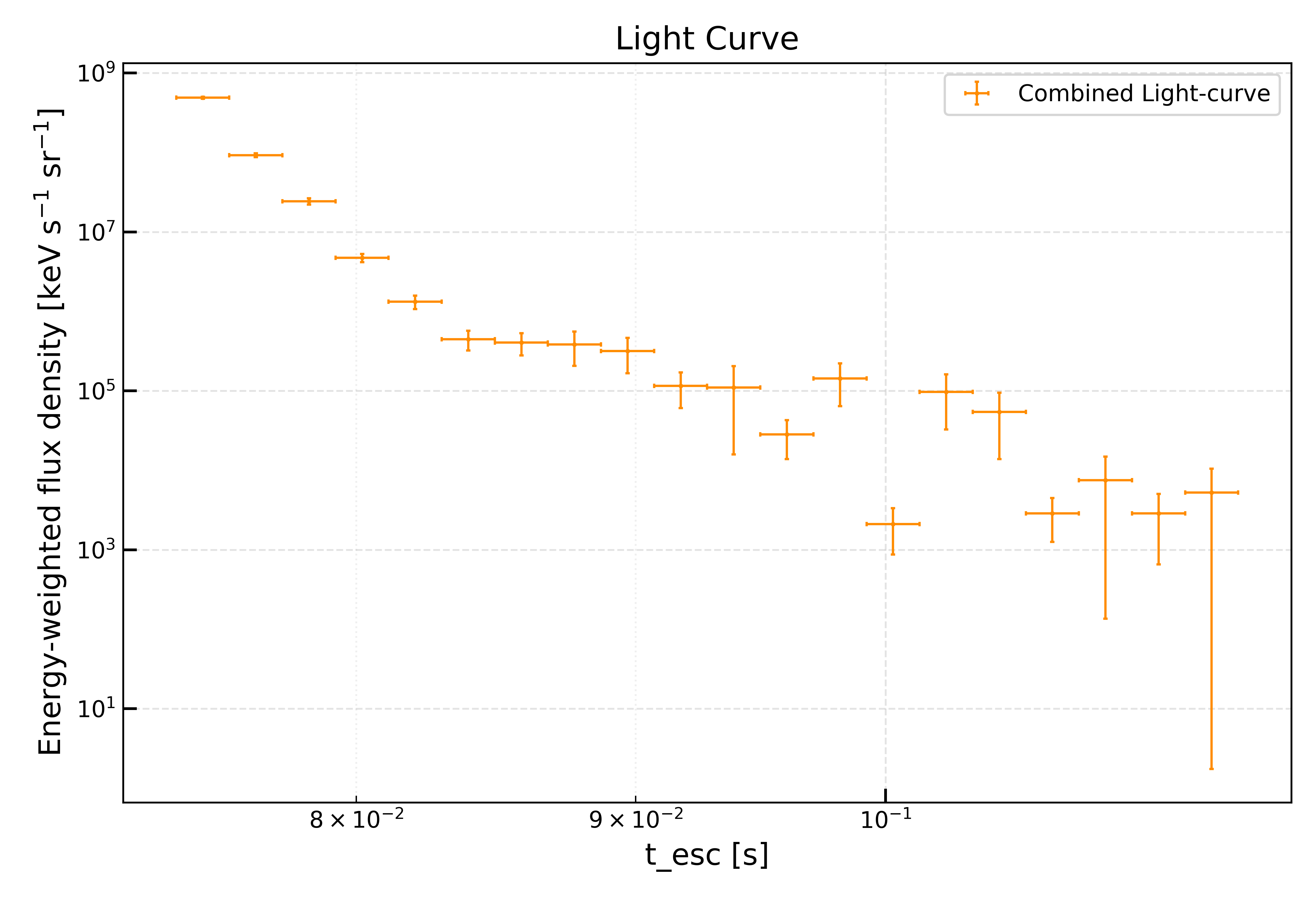}
        \caption{Light curve obtained from instantaneous burst of inner disk photons.}
        \label{fig:light_curve}
    \end{subfigure}
    \hfill
    \begin{subfigure}[b]{0.45\linewidth}
        \centering
        \includegraphics[width=\linewidth]{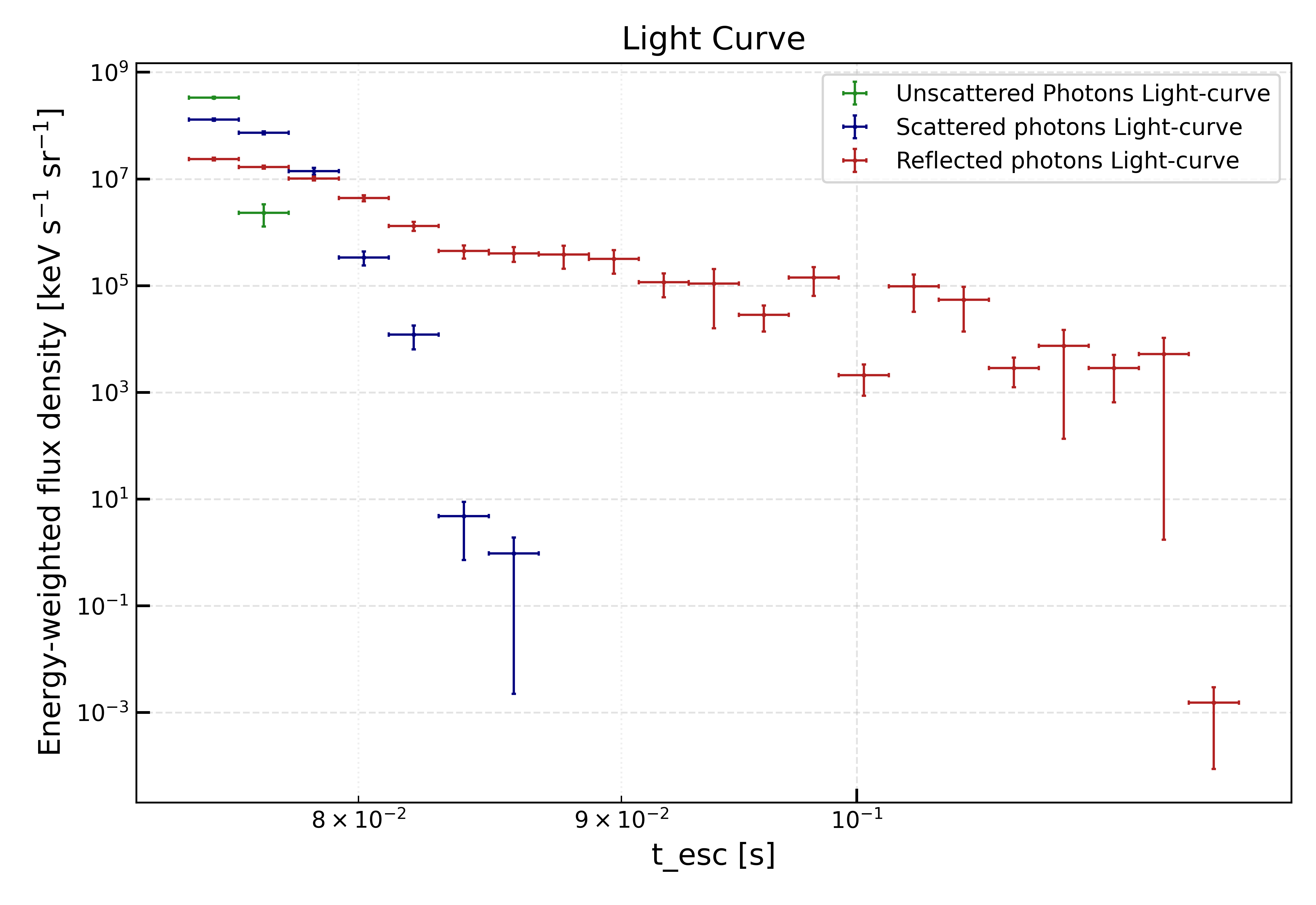}
        \caption{Composition of the light curve with light bending on.}
        \label{fig:light_curve_components}
    \end{subfigure}
    \caption{Photon light-curve distribution and its composition for an instantaneous burst of photons from the inner disc, as seen by an observer at an inclination of $30^{\circ}$ with respect to the BH spin axis and an azimuthal angle of $90^{\circ}$ ($\Delta\theta=\pm2^{\circ}$, $\Delta\phi=\pm2^{\circ}$) relative to the X-axis of the BH system.}
    \label{fig:photon_light_curve}
\end{figure*}

Considering an instantaneous flash of photons from the inner disc, we obtain the light curve of photons arriving at the observer, shown in Fig.~\ref{fig:light_curve}. This distribution characterizes the arrival-time behaviour of the photons, analogous to their path-length distribution, and provides the basis for understanding the time lags discussed in Section~\ref{sec:rev_lags}. The light curve shown in Fig.~\ref{fig:light_curve_components} is further decomposed into the light curves of unscattered photons, corona-scattered photons, and photons that are both corona-scattered and disc-reflected. Each of the three photon classes exhibits a distinct light curve that can be described approximately by a power-law distribution, with the scattered-photon component showing a possible broken power-law feature. The late-time tail of the distribution is dominated by photons that have undergone at least one disc-scattering (reflection) event.

\begin{figure*}
    \centering
    \begin{subfigure}[b]{0.47\textwidth}
        \centering
        \includegraphics[width=\linewidth]{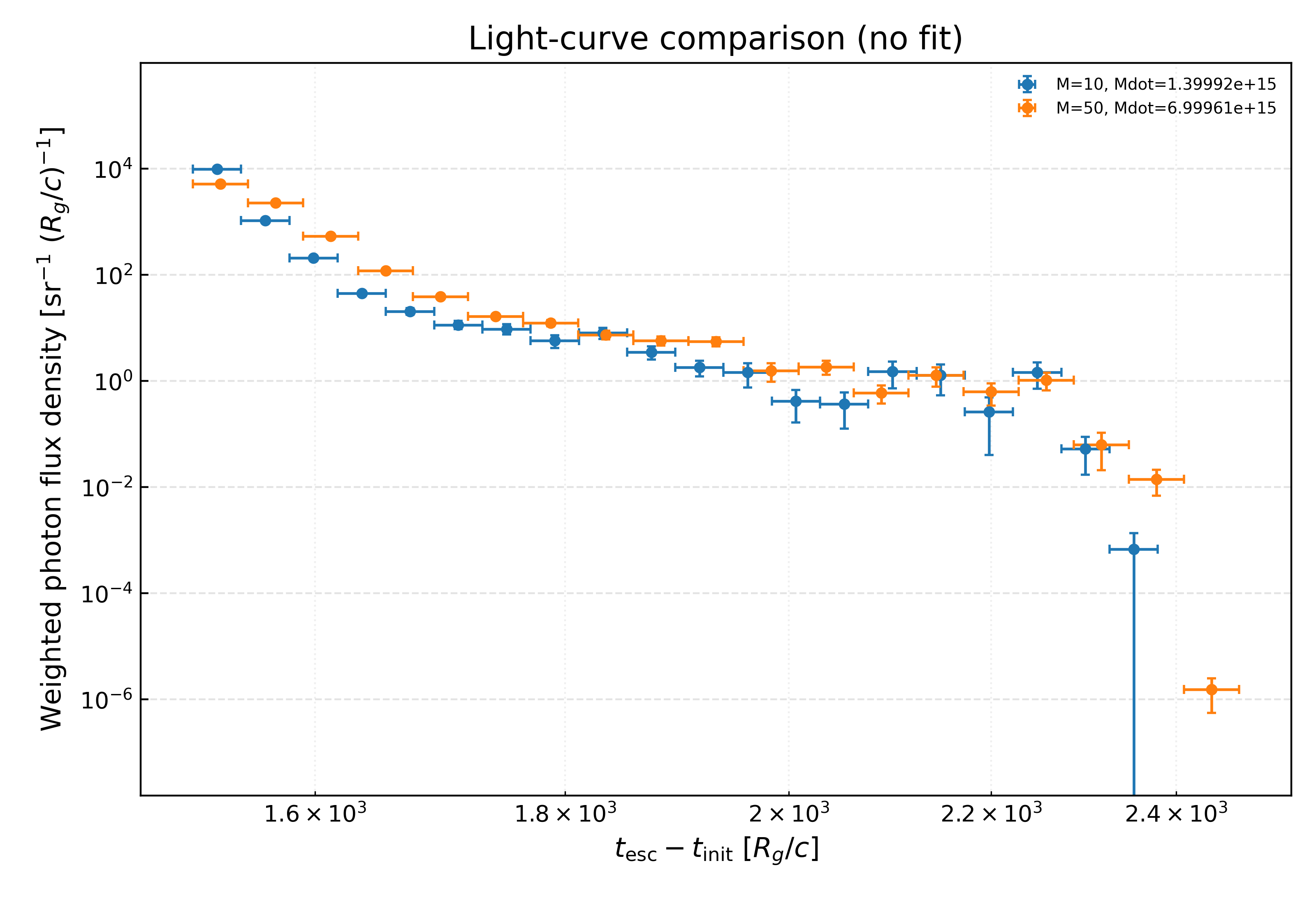}
        \caption{Light curve for varying black hole mass.}
        \label{fig:light_curve_varying_mass}
    \end{subfigure}
    \hfill
    \begin{subfigure}[b]{0.47\textwidth}
        \centering
        \includegraphics[width=\linewidth]{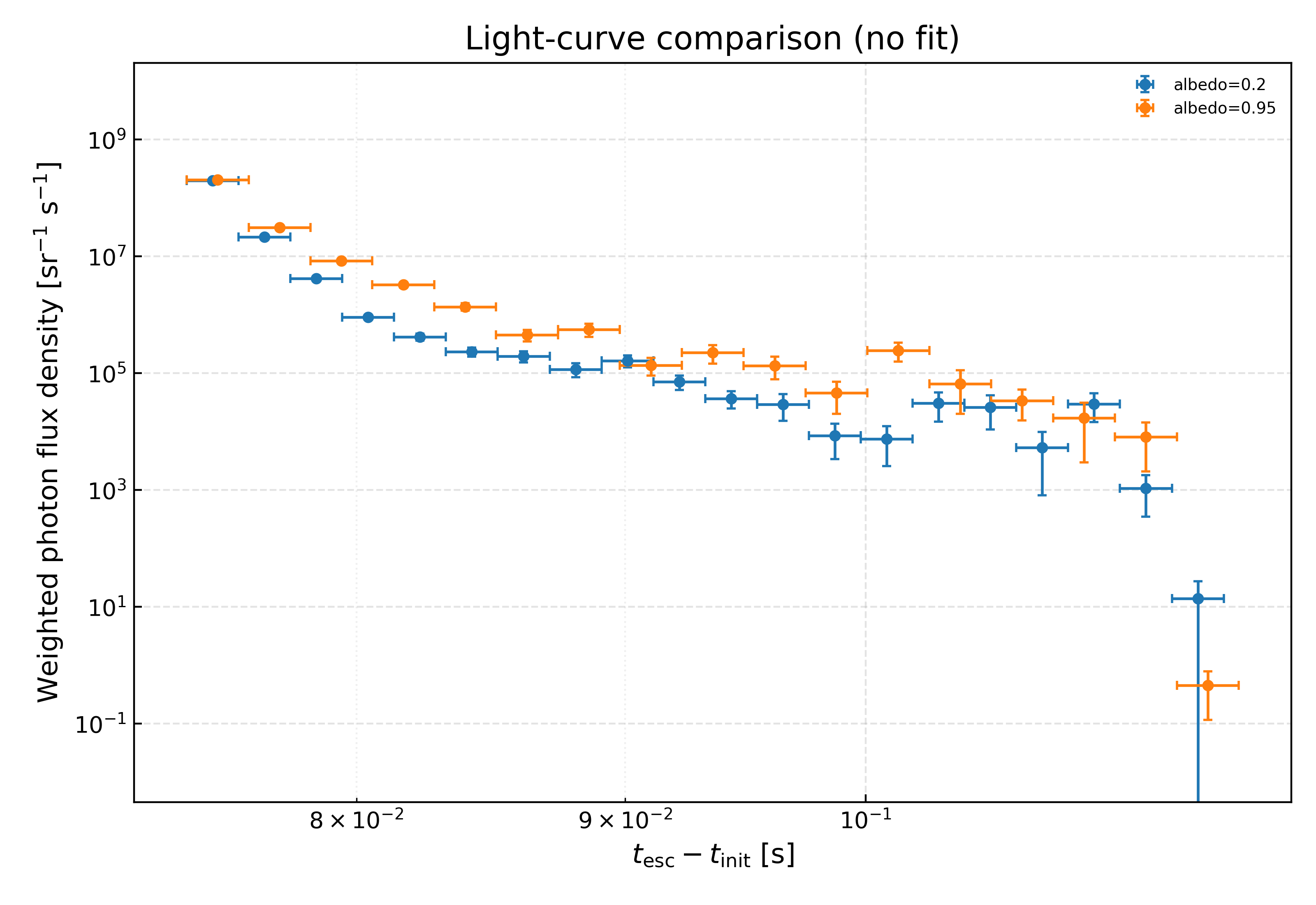}
        \caption{Light curve for varying reflection albedo.}
        \label{fig:light_curve_varying_albedo}
    \end{subfigure}
    \caption{Comparison of simulated light curves under different physical conditions, as seen by an observer at an inclination of $30^{\circ}$ with respect to the BH spin axis and an azimuthal angle of $90^{\circ}$ ($\Delta\theta=\pm3^{\circ}$, $\Delta\phi=\pm3^{\circ}$). The individual panels show the effect of (a) variation in black hole mass and (b) variation in reflection albedo.}
    \label{fig:combined_light_curve}
\end{figure*}

We have also examined the dependence of the light curves on several physical parameters, a subset of which is shown in Fig.~\ref{fig:combined_light_curve}.

\subsection{Spectral lags}
\label{sec:rev_lags}
\begin{figure}
    \centering
    \includegraphics[width=\linewidth]{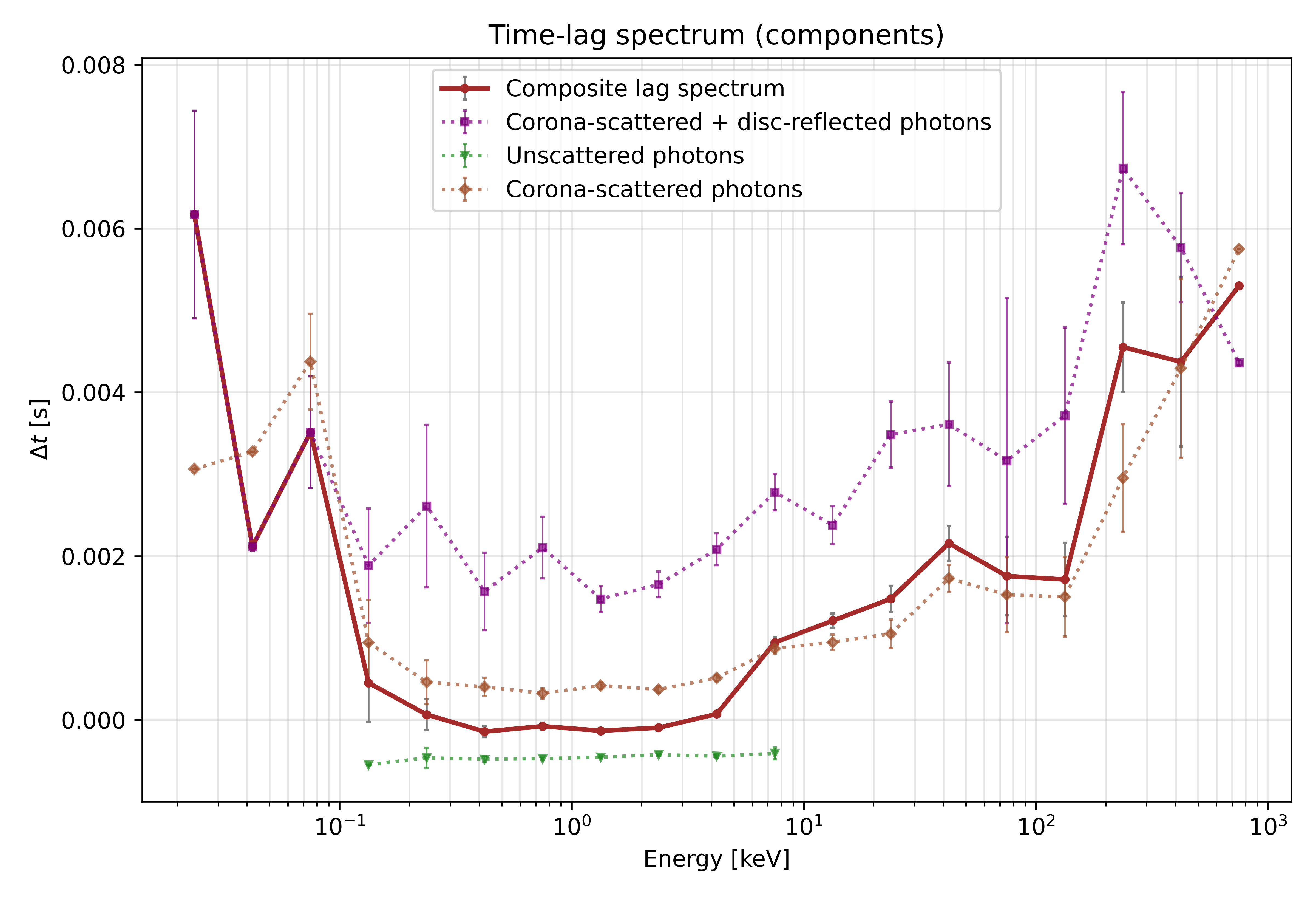}
    \caption{Composite lag spectrum of the photons, decomposed into unscattered, corona-scattered, and corona-scattered plus disc-scattered (reflected) components, as seen by an observer ($\Delta\theta=\pm3^{\circ}$, $\Delta\phi=\pm3^{\circ}$) at an inclination of $30^{\circ}$ with respect to the BH spin axis and an azimuthal angle of $90^{\circ}$.}
    \label{fig:omposit_timelag_spectrum}
\end{figure}
This section examines the composite time-lag spectrum of the system, shown in Fig.~\ref{fig:omposit_timelag_spectrum}. The spectrum is further decomposed into its constituent components. We find that the higher energies are dominated by corona-scattered photons, whereas the lower energies are dominated by disc-scattered (reflected) photons. This behaviour is strongly position dependent, owing to the asymmetry of the system, and at certain positions we also see corona-scattered photons dominating the lower-energy regime over the reflected photons.

\begin{figure*}
    \centering
    \begin{subfigure}{0.33\textwidth}
        \centering
        \includegraphics[width=\linewidth]{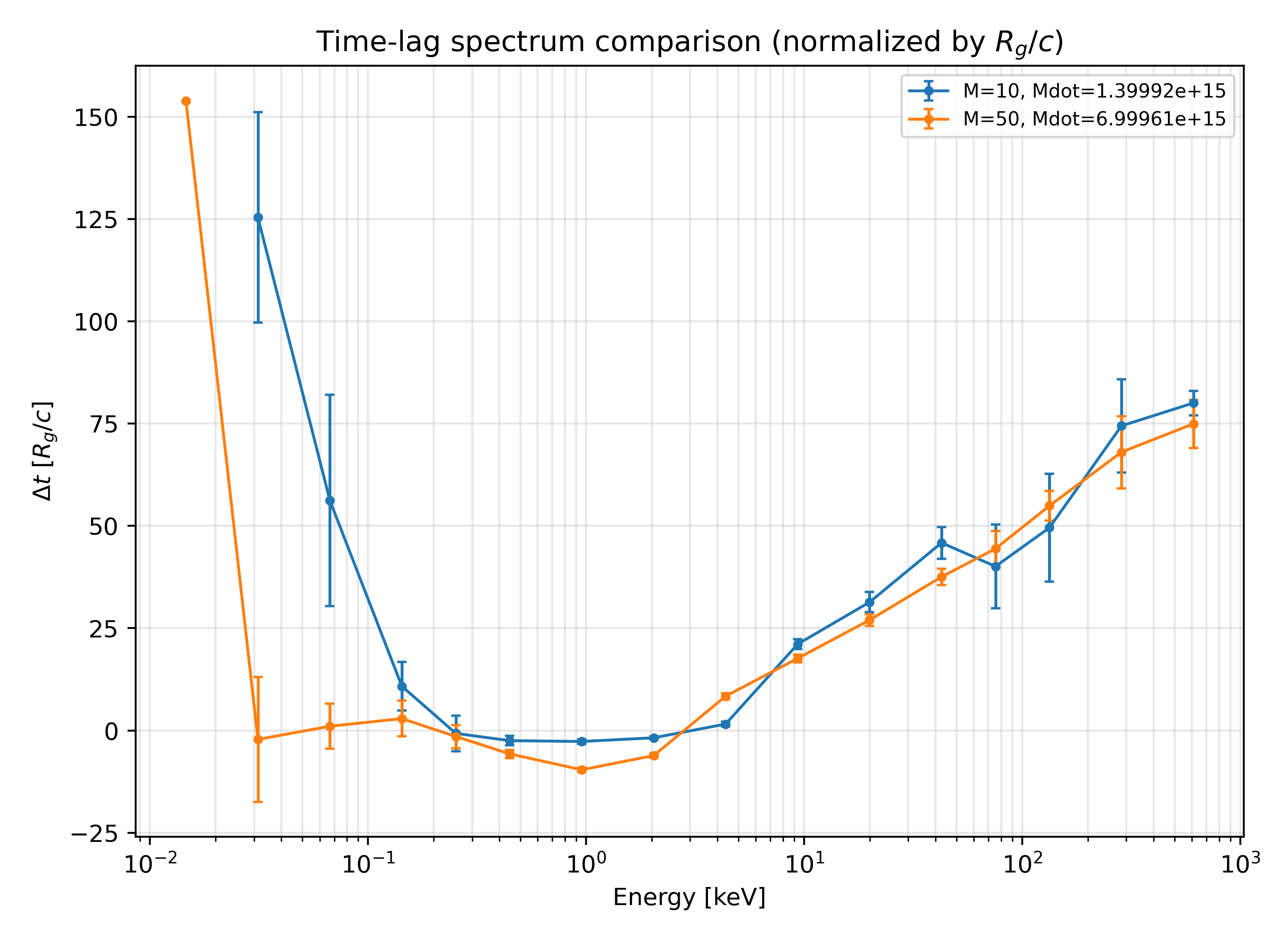}
        \caption{Varying BH mass.}
        \label{fig:lag_spectrum_vs_BH_mass}
    \end{subfigure}
    \hfill
    \begin{subfigure}{0.33\textwidth}
        \centering
        \includegraphics[width=\linewidth]{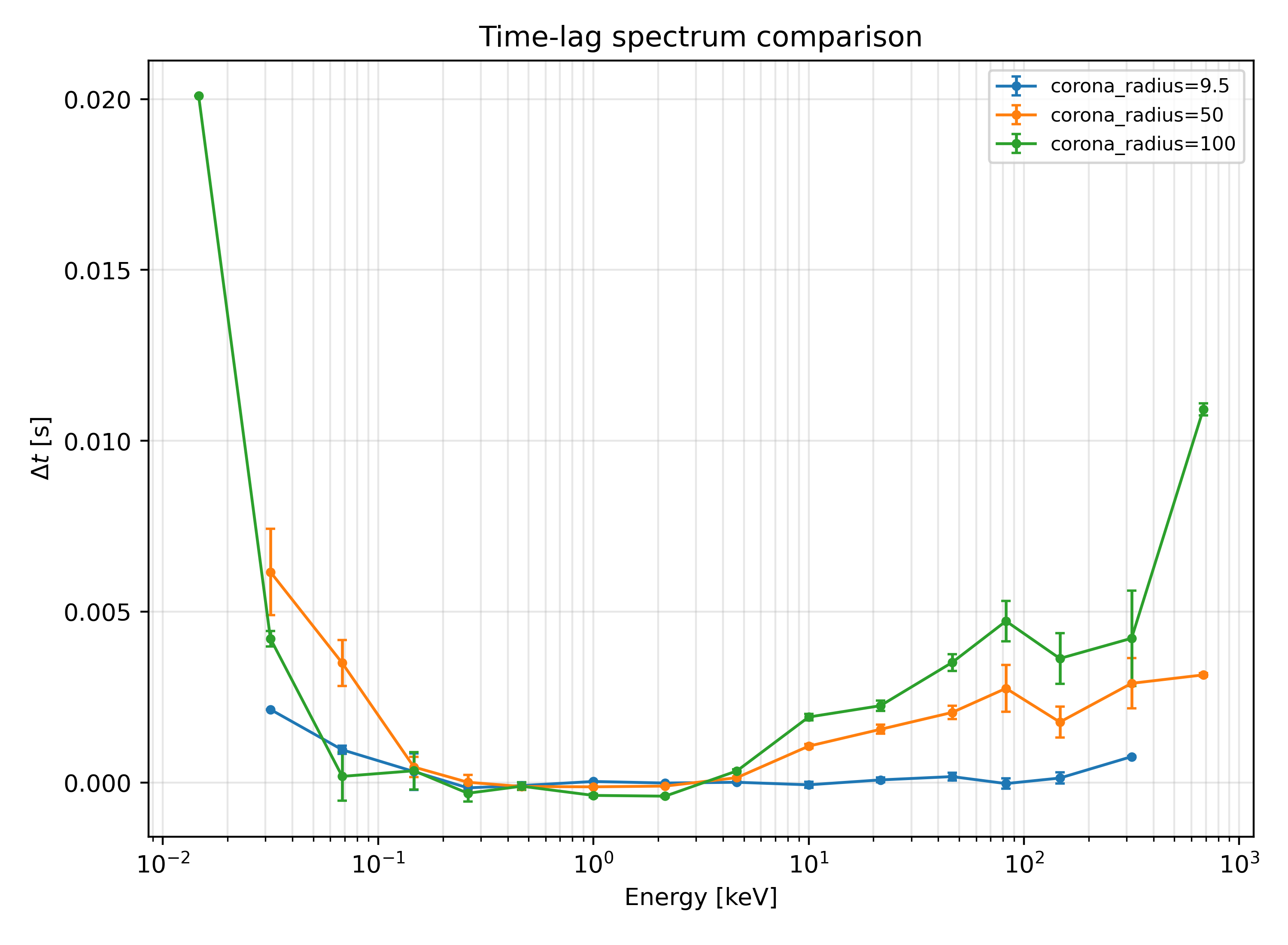}
        \caption{Varying corona radius.}
        \label{fig:lag_spectrum_vs_corona_radius}
    \end{subfigure}
    \hfill
    \begin{subfigure}{0.33\textwidth}
        \centering
        \includegraphics[width=\linewidth]{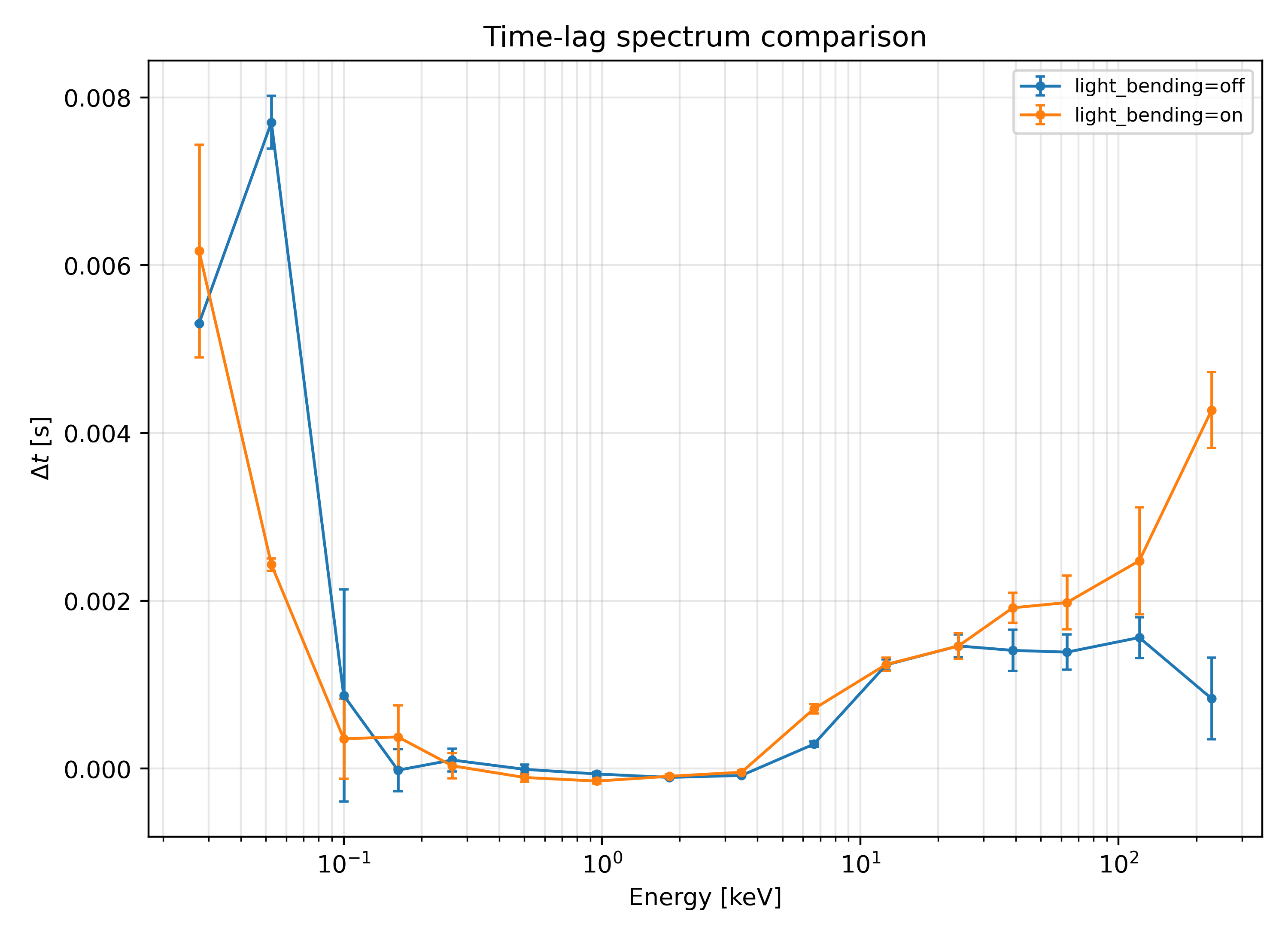}
        \caption{With and without light bending.}
        \label{fig:lag_spectrum_with_and_without_lightbending}
    \end{subfigure}
    \vspace{1em}
    \begin{subfigure}{0.33\textwidth}
        \centering
        \includegraphics[width=\linewidth]{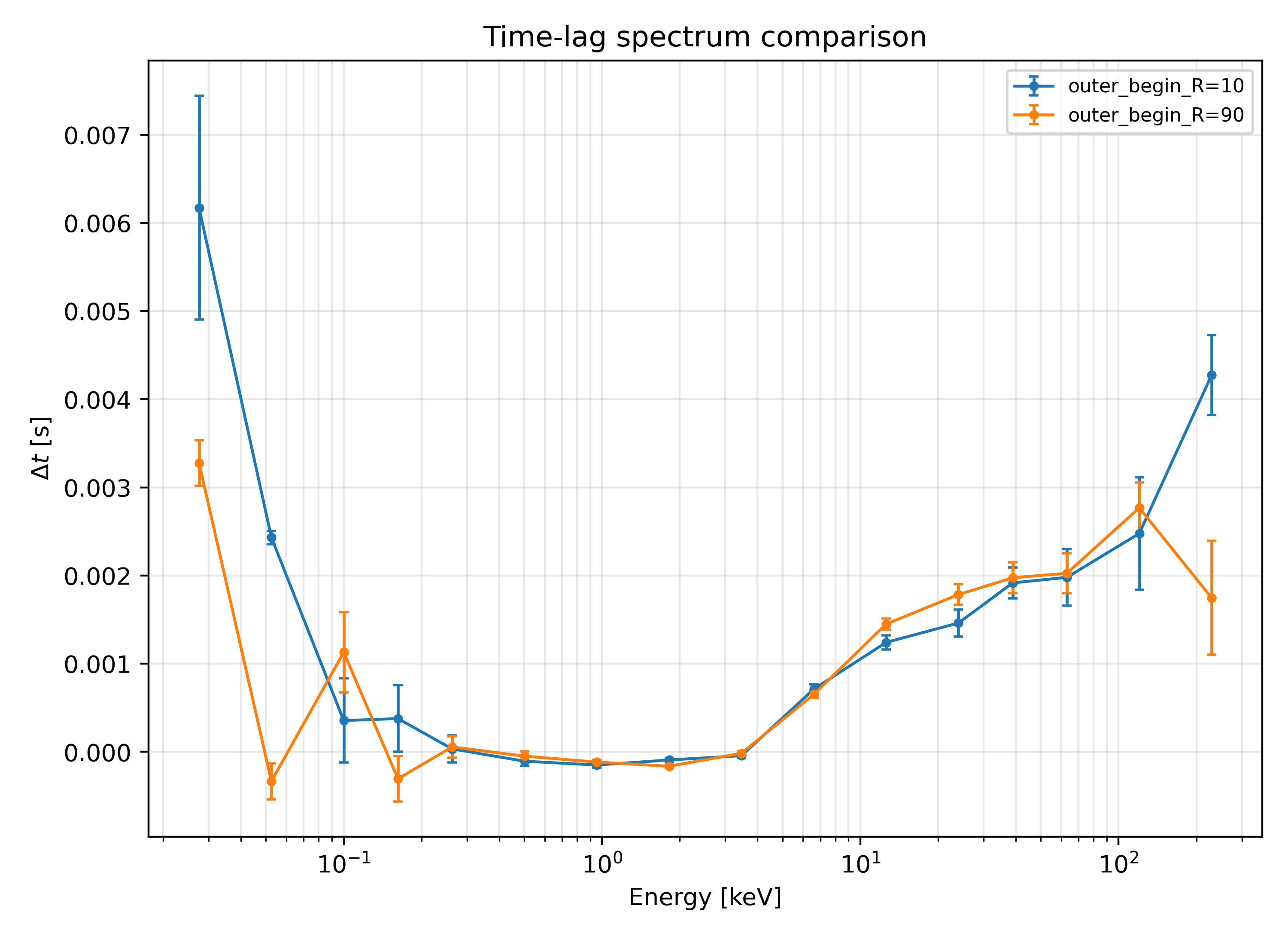}
        \caption{Varying outer-disc inner radius.}
        \label{fig:lag_vs_outer_disk_radius}
    \end{subfigure}
    \hspace{0.08\textwidth}
    \begin{subfigure}{0.33\textwidth}
        \centering
        \includegraphics[width=\linewidth]{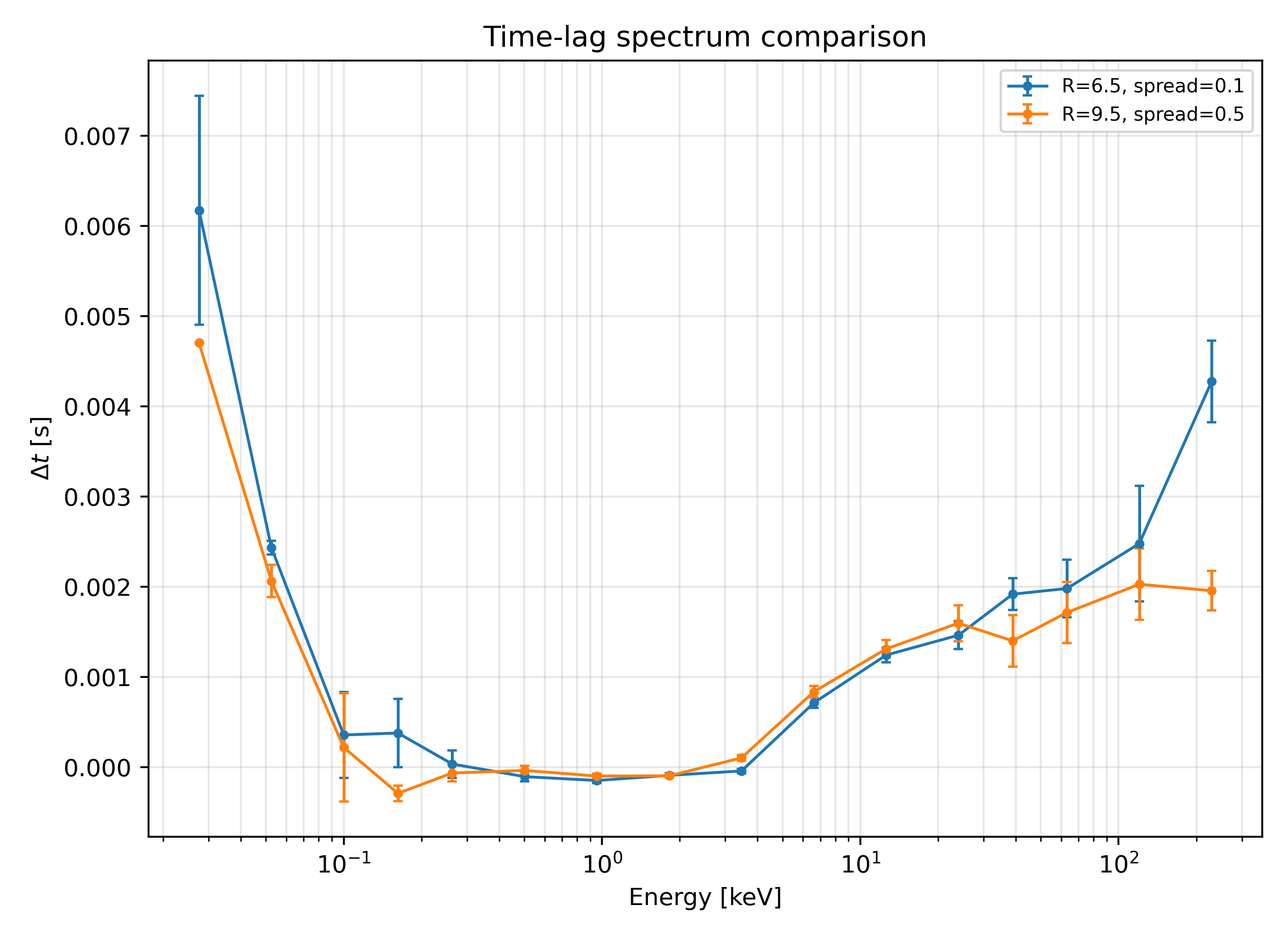}
        \caption{Varying photon-generation region.}
        \label{fig:lag_spectrum_vs_photon_generation_region_width}
    \end{subfigure}
    \caption{Dependence of the time-lag spectrum on various physical system parameters. The individual panels show (a) variation in black hole mass, (b) variation in corona radius, (c) the effect of light bending, (d) variation in the inner boundary of the outer disc, and (e) variation in the photon-generation region, as observed through an observer window of $\Delta\theta=\pm3^{\circ}$, $\Delta\phi=\pm3^{\circ}$, the observer being positioned at an inclination of $30^{\circ}$ w.r.t. the BH spin axis, and having an azimuthal value of $90^{\circ}$ w.r.t. the global X-axis of the system. The R values mentioned in the legend are all in units of $R_g$.}
    \label{fig:lag_spectrum_parameters}
\end{figure*}

We have also examined the dependence of the lag spectrum on various physical system parameters, as well as its behaviour with and without the inclusion of light bending. These results are summarized in Fig.~\ref{fig:lag_spectrum_parameters}.

\subsection{Overall parameter dependence of the simulation}
\label{sec:overall_parameter_dependence}

Based on the energy spectra, time-lag spectra, photon path-length distributions and photon light curves presented in the preceding sections, we conclude that the observed characteristics of the simulated radiation are sensitive to several physical parameters of the model.

These parameters can be broadly grouped into three categories related to the photon propagation and interaction processes. The first concerns \textit{coronal scattering}, including the \textit{number of scattering events within the corona}, the \textit{size of the corona} and the \textit{extent of gravitational light bending}. The second concerns the \textit{outer-disc properties}, particularly its radial extent, characterized by the inner and outer radii ($R_{\rm in}$ and $R_{\rm out}$), and the \textit{reflection probability (albedo)}. The third concerns the \textit{viewing geometry}, primarily determined by the observer inclination, which also influences the observed properties of the emitted and scattered radiation.

\section{Conclusions}
\label{sec:conclusions}

We have presented a general-relativistic Monte Carlo (GRMC) radiative-transfer model of a misaligned, warped, two-zone accretion disc system
around a Kerr black hole, combining a hot spherical Comptonizing corona
around a precessable inner disc with a stationary, reflecting outer disc
truncated at its own Bardeen--Petterson warp radius. By tracking
individual photon histories -- from emission near the ISCO, through
coronal scattering and possible reflection off the tilted outer disc, to
detection by a distant observer -- the model self-consistently links the
composite energy spectrum, photon path-length and arrival-time
distributions, and X-ray spectral time lags within one geometric and
physical picture.

The composite energy spectrum carries a clear imprint of the two-zone
geometry: the low-energy band is dominated by the multicolour blackbody
emission of the outer disc, while the high-energy band is dominated by
Compton up-scattered photons from the inner disc and corona. Decomposing
the inner-disc spectrum further shows that outer-disc-reflected photons
dominate its lower-energy portion, while directly Compton up-scattered
photons dominate its higher-energy portion. The spectral shape and photon
index respond systematically to the black hole mass, corona size and
radius, photon-injection radius, outer-disc geometry and reflection
albedo.

Gravitational light bending leaves a distinct, quantifiable signature on
the photon path-length and arrival-time distributions: switching on light
bending lengthens the tail of the path-length distribution and delays the
arrival of the first photons, which we attribute to curved photon trajectories increasing
the number of scattering events a photon undergoes before escaping. The
resulting light curves decompose cleanly into unscattered, corona-scattered
and corona-scattered-plus-disc-reflected populations, each following an
approximately power-law distribution, with the late-time tail dominated by
at-least-once-reflected photons. The composite lag spectrum
inherits this structure: corona-scattered photons dominate the lag at
higher energies, while disc-reflected photons dominate at lower energies,
and the amplitude of the lag responds systematically to the black hole
mass, corona radius, outer-disc geometry, reflection albedo, and the
presence or absence of light bending.

These results establish the observable, energy- and parameter-dependent
spectral and timing signatures of a tilted, two-zone accretion geometry
around a Kerr black hole, independent of any assumption about photon
polarization. They provide the common radiative-transfer foundation on
which Paper II builds: the same transport engine, extended to carry and
propagate the full photon polarization state via the Walker--Penrose
formalism and a polarized treatment of Compton and disc scattering,
is used there to compute the energy- and precession-phase-resolved
polarization degree and angle of the composite system, and to compare
these predictions directly against IXPE observations of BHXRBs and the
anticipated sensitivity of the upcoming eXTP mission.

As with any model, several simplifications point to directions for future refinement. The disc is treated as two rigid, sharply truncated zones rather than as a continuously warped structure, while the corona is modelled as a static, spherically symmetric region rather than one informed by the density and velocity structure of GRMHD simulations of tilted flows. Similarly, the outer-disc interaction is treated using a Thomson-scattering albedo prescription rather than a full ionized-reflection model, and both the coronal and disc-scattering modules treat Compton scattering in the Thomson limit, using the classical, energy-independent differential cross-section rather than the full Klein--Nishina cross-section. This approximation is justified because only a small fraction of the seed and scattered photons reaching the observer are expected to approach energies comparable to $m_{\rm e}c^2$, while the fraction of electrons populating the high-energy tail of the Maxwell--J\"{u}ttner distribution is likewise small at the coronal temperatures adopted here. Klein--Nishina corrections are therefore expected to affect primarily a small high-energy tail of the predicted spectra. Nevertheless, a fully relativistic Klein--Nishina treatment remains a natural refinement, particularly for harder X-ray energies or higher coronal temperatures. Other natural extensions include adopting a smoother warp profile, implementing a fully energy-dependent Compton redistribution matrix, incorporating GRMHD-informed coronal and disc density and velocity structures, modelling the fluorescent iron-line complex in the reflected spectrum, extending the calculations to slab and lamp-post corona geometries, incorporating the radial variation of electron density along each photon's path rather than approximating it as constant, and accounting for the light-crossing delay associated with the outer disc's own direct emission in the reported spectral lag. These refinements are natural next steps for both papers in this series.

\section*{Acknowledgements}

MC and DB thank Ashoka University and Axis Bank for funding this research via the Axis Bank PhD grant. The simulations in this paper were run on three computing facilities at Ashoka University: the Gsharp cluster (courtesy of the Physics Department) and Tao (courtesy of Prof.\ Garima Mishra, with access facilitated by Souradeep Sengupta and Bidisha Mukherjee), along with one computational facility of Prof. Suratna Das, (with access facilitated by Debadyuiti Ghosh). We also acknowledge the use of AI tools (Claude, ChatGPT and Gemini) for assistance with writing and text-level editing.

\section*{Code Availability}

The simulation code underlying this article may be shared on
reasonable request to the corresponding author.

\bsp
\label{lastpage}
\end{document}